\documentclass[aps,prd,twocolumn,groupedaddress,showpacs,nofootinbib]{revtex4}
\usepackage{graphicx,dcolumn,bm,amssymb,amsmath,latexsym,amsfonts,footnote}
\usepackage[usenames]{color} %quitar al final este comando
\usepackage{float}
\begin{document}

%%%%%%%%%%%%%%%%%%%%%%%%%%%%%%%%%%%%%
%defining some commands
\newcommand{\nonu}{\nonumber}
\newcommand{\sm}{\small}
\newcommand{\noi}{\noindent}
\newcommand{\npg}{\newpage}
\newcommand{\nl}{\newline}
\newcommand{\bp}{\begin{picture}}
\newcommand{\ep}{\end{picture}}
\newcommand{\bc}{\begin{center}}
\newcommand{\ec}{\end{center}}
\newcommand{\be}{\begin{equation}}
\newcommand{\ee}{\end{equation}}
\newcommand{\beal}{\begin{align}}
\newcommand{\eeal}{\end{align}}
\newcommand{\bea}{\begin{eqnarray}}
\newcommand{\eea}{\end{eqnarray}}
\newcommand{\bnabla}{\mbox{\boldmath $\nabla$}}
\newcommand{\univec}{\textbf{a}}
\newcommand{\VectorA}{\textbf{A}}
\newcommand{\Pint}
%%%%%%%%%%%%%%%%%%%%%%%%%%%%%%%%%%%%

\title{Energy extraction from the Reissner-Nordstr\"om de Sitter black hole}

\author{A. Baez$^{1,}$\footnote{jose.baez@cinvestav.mx}, Nora Breton  $^{1,}$\footnote{nora.breton@cinvestav.mx}, and I. Cabrera-Munguia $^{2,}$\footnote{icabreramunguia@gmail.com}}
\affiliation{$^{1}$ Departamento de F\'isica, Centro de Investigaci\'on y de Estudios
 Avanzados del Instituto Politecnico Nacional; Apdo. Postal 14-740, Mexico City, Mexico\\
$^{2}$Departamento de F\'isica y Matem\'aticas, \\
Universidad Aut\'onoma de Ciudad Ju\'arez, 32310 Ciudad Ju\'arez, Chihuahua, M\'exico}

%\date{\today}

%------------------------------Begin of the document --------------------------------

\begin{abstract}
The energy extraction from an electrostatic black hole  by the decay or splitting of electrically charged particles  is analyzed. We determine the energetic conditions that make the extraction process viable and present a general expression for the efficiency in terms of the parameters of the electrostatic black hole and the decaying particles. We also examine the conditions that optimize the efficiency of the extraction process.  We analyze two particular cases, the first one is the extraction process from a Reissner Nordstr\"om black hole, for charged test particles with nonvanishing angular momentum; the second one and more interesting corresponds to the energy extraction from a Reissner Nordstr\"om de Sitter black hole. For the latter there are two regions where the energy extraction is possible, the generalized ergosphere and a cosmological ergosphere induced by the cosmological horizon. Under certain conditions the two ergospheres get connected and cover the whole region between the event horizon and the cosmological horizon, and therefore the energy extraction is possible at any point in the vicinity of the black hole. Moreover, the efficiency of the energy extraction can be the same for different break up points and also there is the possibility of a different efficiency for the same break up point.  The conditions  that maximize the efficiency are determined as well.
\end{abstract}
%\pacs{04.20.Jb, 04.70.Bw, 97.60.Lf}

\maketitle

\section{Introduction}
The Penrose process is a mechanism proposed in \cite{PenFloyd1971}  to  extract energy from a rotating black hole (BH) like the Kerr spacetime \cite{Kerr1963,Visser2007}; it considers a decay or splitting process in the region delimited between the event horizon and the stationary limit surface. When a particle gets in the ergosphere a timelike trajectory becomes spacelike and the particles can have negative energy (as measured by an observer at infinity); a particle inside  the ergosphere still can  avoid enter into the event horizon and can escape back to infinity. \\
Essentially the Penrose process consists of an ingoing particle that falls freely into the BH; the particle penetrates the ergosphere and then it decays into two fragments, one them with negative energy that remains confined within the ergosphere and eventually it falls into the event horizon. Due to energy conservation, if the second fragment escapes to infinity, it does with a greater energy than the original particle.\\
 Because the energy extraction process takes place inside the ergosphere, that is a characteristic region of stationary solutions, in principle, the extraction process is not possible  from static BHs. Nevertheless, as a variation of Penrose process is the electric Penrose process\ \cite{Christo1971,DenardoRuffini1972,Dadich1980} that allows the energy  extraction  from electrically charged BH, even for static BHs, via electrostatic interaction between the charged black hole and charged test particles. Recent studies about electric Penrose process include the analysis of energy extraction from a Reissner-Nordstr\"om-anti-de Sitter BH\ \cite{Lemos2024} and the possibility to obtain BH energy factories and BH bombs is studied considering a recursive Penrose process; other works that carry out the energy extraction from static BHs are\ \cite{Richartz2021, Baez2022},  where systems of equally and oppositely charged BHs  described by Majumdar-Papapetrou\ \cite{Majumdar,Papapetrou,Hartle1972} and Bonnor\ \cite{Bonnor1979,Cabrera2011} metric,  are analyzed.\\
There are other variations of the Penrose process that can be associated with some astrophysical observations. For instance, collision in the Penrose process might eventually eliminate dark energy particles in the vicinity of a supermassive BH once the multiple particles that scatter inside the ergosphere achieve an arbitrarily high center of mass energy \cite{Schnittman2018, BSW2009}. On the other hand, the presence of an external magnetic field surrounding a rotating BH can give rise to accretion disks comprising charged ionized matter\ \cite{Kolos2021}, this could potentially be connected to the observation of high-frequency oscillations observed in microquasars or galactic nuclei where ultra high-energy particles around rotating magnetized BHs are created \cite{Kolos2017, Kolos2015}. The radiative Penrose process is related to synchroton radiation of charged particles interacting in the ergosphere of a magnetized BH, where the process could produce a particular type of photons having negative energy relative to a distant observer \cite{Kolos2021, Kolos2018}. On the other hand, the electromagnetic Penrose process  \cite{Bhat1985,Tursunov2021,Parthasarathy1986,Wagh1985,Wagh1989,Nucamendi2022} allows events of high energy emission that improve significantly the efficiency of  20\% of the Penrose process due solely to the rotation \cite{Bardeen1972,Wald1974}. It is worth mentioning, that exist other ways of extracting energy from a BH, for instance, the superradiance that considers waves instead particles. In superradiance a wave field is sent into the ergoregion, and under the appropriate conditions an amplified wave returns to infinity. The superradiant effects are exhibited by Kerr BH \cite{Starobinsky1973,Teu1974}.
We also point out that the process here analyzed is different from the energy extraction by Hawking radiation, mainly because it does not require a time varying background geometry, and the whole treatment is classical \cite{Damour75}.

The present paper aims to investigate energy extraction from a black hole via the electric Penrose process. In \cite{DenardoRuffini1972} was analyzed for the first time the extraction from a Reissner-Nordstr\"om BH via electrostatic interaction considering the decay of a charged particle, however the study was restricted to zero angular momentum of the decaying particle. In \cite{Lemos2024} the recursive Penrose process for a Reissner-Nordstr\"om-anti-de Sitter BH is studied. Even analysis on binary solutions of electrostatic BHs  has been developed\ \cite{Richartz2021, Baez2022}. We present the analysis of the electric Penrose process for a static, spherically symmetric charged BH; we derive a general expression for the efficiency of the energy extraction and determine the energetic conditions that optimize the efficiency. We extend the study of the RN BH presented in \cite{DenardoRuffini1972} to arbitrary angular momentum of the charges involved; additionally, the particle parameters that maximize the efficiency of the process are determined. The second case that is examined is the Reissner-Nordstr\"om-de Sitter BH;  we have determined the regions where extraction energy is possible and its relationship with the charges of the test particles.\\
Our paper is organized as follows. In Sec.\ref{sec21}, we introduce the general static  spherically symmetric spacetime and derive the motion equations for charged massive test particles. In Sec.\ \ref{secergo}  the generalized ergosphere and  a general expression to determine the efficiency are presented; the test particle conditions that optimize the process are discussed as well. In Sec. \ref{RNextr} it is analyzed the Penrose process for the RN BH, in the general case of nonvanishing angular momentum of the particle with negative energy, obtaining in the limit of zero angular momentum the results in \cite{DenardoRuffini1972}.
In Sec.\ \ref{secRNLamnda} the analysis of the extraction process (efficiency, generalized ergosphere and restrictions) for Reissner-Nordstr\"om-de Sitter is presented; the RNdS BH has two ergospheres and there are cases with the same  efficiency for different break up points and also the possibility of a different efficiency for the same break up point. In the last section Conclusions are given.

\section{Penrose mechanism for electrostatic black holes}
\vspace{-0.3cm}
Considering the Penrose process introduced in \cite{PenFloyd1971}, we present an approach to study the efficiency of this process in the vicinity of an electrostatic BH. It consists in sending a charged particle towards the BH; at some point, inside the  ergosphere, the particle breaks up into two fragments, one of them escapes to infinity with more energy than the initial one, while the second fragment remains inside the ergosphere and eventually penetrates the horizon of the BH. This mechanism requires that the BH possesses an ergoregion, which, in principle, only stationary BH possess; therefore by means of the regular Penrose process it is not possible to extract energy from a static BH. However, for electrostatic BH it is possible to define a  generalized ergosphere where charged test particles can have negative energies \cite{Christo1971,DenardoRuffini1972,Dadich1980}. 

Let us consider the static spherically symmetric line element,
\begin{equation}\label{smetric}
ds^2 = g_{tt} dt^2 +g_{rr} dr^2 + g_{\theta\theta}d\theta^2 + g_{\phi\phi}d\phi^2,
\end{equation}
\noi where $(t, r,\theta,\phi)$ are spherical-like coordinates and the components of the metric $g_{\mu\nu}$ only depend of $r$ coordinate. We use the signature $(-,+,+,+)$. 
The electric potential $A_{\mu}$ of the charged BH is 

\begin{equation}
A_{\mu}=(A_t, 0,0,0),
\end{equation}
%%%%%%%%%%%%%%%%%%%%%%%%%%%%%%%%%%%%%%%%%
\subsection{Motion of charged particles}\label{sec21}

The  equation of motion for a test particle with charge $q$,  mass $m$, and charge mass ratio  $e=q/m$,  can be obtained from the Euler-Lagrange equations with the Lagrangian 
\begin{equation}
\mathcal{L} = \frac{1}{2} g_{\mu\nu}\dot{x}^{\mu}\dot{x}^{\nu} + e A_{\alpha}\dot{x}^{\alpha},
\end{equation}
\noi where the dot means derivative with respect to an affine parameter $\dot{x}^\mu =dx^{\mu}/d\lambda$. In the case of a massless particle $\lambda$ is a properly chosen affine parameter. For a particle of mass $m$ the affine parameter is $\lambda=\tau/m$, where $\tau$ is the proper time. In  terms of the metric coefficients in Eq. (\ref{smetric}), the Lagrangian is
\begin{equation}
\mathcal{L} = \frac{1}{2}\left(g_{tt}\dot{t}^2+g_{rr}\dot{r}^2+g_{\theta\theta}\dot{\theta}^2+g_{\phi\phi}\dot{\phi}^2\right)+e A_{t}\dot{t},
\end{equation}

\noi that does not depend explicitly on the $(t,\phi)$ coordinates. Then we can identify two motion constants of the test particle: its energy and angular momentum per unit mass  $\mathcal{E}$ and $l$ respectively, given by
\begin{equation}
l=\frac{L}{m}=\frac{\partial\mathcal{L}}{\partial\dot{\phi}}=g_{\phi\phi}\dot{\phi},\quad
\mathcal{E}=\frac{E}{m}=-\frac{\partial\mathcal{L}}{\partial\dot{t}}=-g_{tt}\dot{t}-eA_{t},
\end{equation}
where $E$ and $L$ are the energy and angular momentum, respectively. The  motion equations are obtained after solving for $\dot{\phi}$ and $\dot{t}$, 
\begin{equation}
\dot{\phi}=\frac{l}{g_{\phi\phi}},\quad \dot{t}=-\frac{\mathcal{E} +eA_{t}}{g_{tt}}.    
\end{equation}

\noi and plugging them into the contraction of the four moment $\dot{x}^\mu \dot{x}_{\mu}=-\delta$ ($\delta=1$ for massive particles and $\delta=0$ for massless particles), obtaining
\begin{equation}\label{geodesics}
\begin{aligned}
\dot{r}^2 =&-\frac{1}{g_{rr}}\left(\frac{\left(\mathcal{E}+eA_t\right)^2}{g_{tt}}+\frac{l^2}{g_{\phi\phi}}+\delta\right).
\end{aligned}
\end{equation}
Due to the spherical symmetry, without loss of generality, we restrict ourselves to the motion on the equatorial plane, then $\theta=\pi/2$ and $\dot{\theta}=0$.

The radial motion has turning points at $\dot{r}=0$, that is a quadratic equation in $\mathcal{E}$, whose solutions are the effective potentials,  $\mathcal{E}=V_{\pm}$, given by
\begin{equation}\label{potefectivo}
V_{\pm}=\pm\sqrt{-\frac{g_{tt}}{g_{\phi\phi}}l^2-g_{tt} \delta}-e A_{t}.
\end{equation}
Note that the effective potential is different for each particle, i.e. the particle ``i" with angular momentum $l_i$ and charge mass ratio $e_i$ moves under the effective potential
\begin{equation}\label{potefectiv_i}
V_{\pm i}=\pm\sqrt{-\frac{g_{tt}}{g_{\phi\phi}}l_{i}^2-g_{tt} }-e_i A_{t}.
\end{equation}
Along the paper we analyze the potential corresponding to the particle that penetrates the horizon, the particle 1.

\subsection{Generalized ergosphere}\label{secergo}
The energy extraction from an electrostatic BH  is possible if negative energy states (NES) exist for a test particle; the NES are possible if  
$\mathcal{E} = V_+ < 0$.

The region that admits NES is called the``generalized ergosphere", it is  the region  delimited by the event horizon $r_h$ and the radius $r_{e}$, defined from the condition $ \mathcal{E} = V_+ = 0$,
\begin{equation}\label{regcondition}
\sqrt{-\frac{g_{tt}}{g_{\phi\phi}}l^2-g_{tt} \delta}-e A_{t}=0,
\end{equation} 
and such that  $r_h < r_{e}$.
For non charged test particles, $e=0$, the condition (\ref{regcondition}) implies $g_{tt}=0$, equivalently,
$r_h = r_{e}$, and there is not ergosphere. Therefore a necessary condition for the existence of NES is that charged test particles interact with the electrostatic BH; i.e. the extracted energy is to the expense of the electrostatic BH energy. 

Additionality, it can be shown that the maximum region admitting NES is obtained when the test particle angular momentum is zero $l=0$, from Eq. (\ref{regcondition}) this leads to the condition,
\begin{equation}\label{regcondition2}
g_{tt}  +  \left(e A_t\right)^2=0.
\end{equation}
%%%%%%%%%%%%%%%%%%%%%%%%%%%%%%%%%%%%%%%%%%%%%%%%%%%%%%%%%%%%%%%%%%%%%%%%%%%%%%%%%%%%%%%%

%%%%%%%%%%%%%%%%%%%%%%%%%%%%%%%%%%%%%%%%%%%%%%%%%%%%%%%%%%%%%%%%%%%%%%%%% 
%\subsection{Efficiency of the Penrose process}\label{pp}

Let us consider a charged test particle moving along a timelike geodesic (particle $0$)  that reaches one turning point ($\dot{r}=0$); at that point it breaks up into two pieces, one of them with negative energy (particle $1$) and the other one with positive energy (particle $2$). The particle with negative energy is confined inside the generalized ergosphere until it falls into the BH, while the other one escapes from the generalized ergosphere with more energy than the initial one. 
The $T_i$ trajectories are timelike paths $x_i^\mu (\lambda)$ parametrized by $\lambda=\tau/m$ where $\tau$ is the proper time;
the incident particle follows the trajectory $T_0$, which starts outside the ergosphere and ends inside it at the break-up point ($r_*$,$\theta_*$,$\phi_*$). From the break-up point emerge two particles with trajectories labeled as $T_1$ for the particle with negative energy ($E_1<0$) that remains in the ergosphere; the trajectory $T_{2}$ corresponds to the particle escaping outside the ergosphere. We denote with $m_i$, $e_i$, $\mathcal{E}_i$, $l_i$ and $p_i^\mu = dx^\mu_i/d\lambda$ to the mass, charge mass ratio, energy and azimuthal angular momentum per unit mass, and  4-momentum of the $i$ particle, respectively. These quantities should fulfill the charge and 4-momentum conservation equations, 
\begin{equation}\label{qconservation}
m_0 e_0=m_1 e_1+m_2 e_2,
\end{equation}
and
\begin{equation}\label{4momentum}
 p_0^\mu=p_1^\mu+p_2^\mu.
\end{equation}
 
 At the break-up point the energy is conserved (time 4-momentum component $p^{t}$); while from the spatial components of the 4-momentum it is derived the conservation of linear momenta, namely,
\begin{align}
m_0 \mathcal{E}_0=m_1 \mathcal{E}_1+m_2 \mathcal{E}_2,\label{Econservation}\\
 m_0 l_0=m_1 l_1+m_2 l_2.\label{Lconservation}  
\end{align}

To determine the efficiency of the Penrose process we need to know the explicit form of the energies $E_i$ of each particle at the break-up point  where $\dot{r}=0$ (turning point). From the radial Eq. (\ref{geodesics}), we obtain the energy and angular momentum for the particle $i$  at the break-up point, given by 
\begin{equation}\label{E0}
\mathcal{E}_i=\sqrt{-\frac{g_{tt}}{g_{\phi\phi}}}\sqrt{l_i^2+g_{\phi\phi}\delta_i}-e_iA_t, \quad i=0,1,2.
\end{equation}
and
\begin{equation}\label{angularmomentum}
l_i=\pm\sqrt{-\frac{g_{\phi\phi}}{g_{tt}}}\sqrt{\left(\mathcal{E}_i+e_i A_t\right)^2+g_{tt}\delta_i} \quad i=0,1,2.
\end{equation}

According to Eq. (\ref{regcondition}), $V_{+}$ and then the NES are independent of the sign of $l$; thus we can choose $l_i$ with any sign. Using Eq. (\ref{angularmomentum}), considering $l_i>0$ and Eq. (\ref{Lconservation}) we obtain,
\begin{equation}\label{angularmomentum2}
\begin{aligned}
&\sqrt{\left(\mathcal{E}_0+e_0 A_t\right)^2+g_{tt}\delta_0}-\sqrt{\left(\mathcal{E}_1+e_1 A_t\right)^2+g_{tt}\delta_1}\\
&\qquad\qquad-\sqrt{\left(\mathcal{E}_2+e_2 A_t\right)^2+g_{tt}\delta_2}=0;
\end{aligned}
\end{equation}
solving Eq. (\ref{angularmomentum2}) and considering Eq. (\ref{Econservation}) we obtain

\begin{equation}\label{energies}
\begin{aligned}
&\tilde{E}_1=\frac{1}{2} \left(\left(1+\frac{\tilde{\delta}_1}{\tilde{\delta}_0}-\frac{\tilde{\delta}_2}{\tilde{\delta}_0}\right)\tilde{E}_0+\kappa\sqrt{d_0}\right),\\
&\tilde{E}_2=\frac{1}{2}\left(\left(1-\frac{\tilde{\delta}_1}{\tilde{\delta}_0}+\frac{\tilde{\delta}_2}{\tilde{\delta}_0}\right)\tilde{E}_0-\kappa\sqrt{d_0}\right),\\
& \tilde{E}_i=E_i+q_i A_t, \quad i=1,2, \quad \kappa=\pm 1,\\
d_0=& \left( 1- 2 \left( \frac{\tilde{\delta}_1+\tilde{\delta}_2}{\tilde{\delta}_0}\right) + \left( \frac{\tilde{\delta}_1-\tilde{\delta}_2}{\tilde{\delta}_0} \right)^2 \right)
\left(\tilde{E}_0^2+g_{tt}\tilde{\delta}_0 \right) .
\end{aligned}
\end{equation}

%%%%%%%%%%%%%%%%%%%%%%%%%%%%%%%%%%%%%%%%%%%%%%%%%%%%%%%%%%%%%%%%%

%%%%%%%%%%%%%%%%%%%%%%%%%%%%%%%%%%%%%%%%%%%%%%%%%%%%%%%%%%%%%%%%%%
\noi where $E_i=m_i \mathcal{E}$, $q_i=m_i e_i$ are the energy and charge of the $i$-particle, and $\tilde{\delta}_i=m^2_i \delta_i$, $\tilde{\delta}_i$ is $m^2_i$ for massive particles and zero for massless particles. It is important to remark that the parameter that defines  the generalized ergosphere is the charge-mass ratio of the particle that penetrates the BH, $e_1=q_{1}/m_{1}$, while the generalized ergosphere is independent on the charge and mass of the ingoing and outgoing particles, $0$ and $2$. 
The results in \cite{DenardoRuffini1972} for the RN BH are recovered with $\kappa=-1$. We just noticed that in \cite{Zaslavski2024} the RN BH energy extraction is analyzed considering an arbitrary break up point, i.e., not restricted to be a turning point.

\subsection{Efficiency of the electric Penrose process}

\noi The efficiency $\eta$ of the Penrose process can be defined as the ratio between the gain in energy  and the input energy (energy of the incident particle), 
\begin{equation}\label{eff1}
\eta=\frac{E_2-E_0}{E_0}=-\frac{E_1}{E_0}.
\end{equation}

\noi where the energies $E_2$ and $E_1$ are given in Eq. (\ref{energies}). Note that $E_0$ implicitly depends on the break-up point  $r_*$ and on the angular momentum $L_0$,  that depends on $r_*$ as well.

To obtain a general expression for the efficiency,  $\eta$,  for an arbitrary break-up point $r_*$,  we need to know the energies $E_{0}$, $E_{1}$ and $E_{2}$ in terms of  $r_*$. The radial equation, Eq. (\ref{geodesics}), for a turning point $\dot{r}=0$, leads to the system of equations,
\begin{equation}\label{eqenergies}
g_{\phi\phi}\left(\mathcal{E}_i+e_i A_{t}\right)^2+g_{tt}\left(l_i^2+g_{\phi\phi}\delta_i\right)=0, \quad i=0,1,2.
\end{equation}
Substituting the conservation of charge, energy and angular momentum of Eqs. (\ref{qconservation}), (\ref{Econservation}) and (\ref{Lconservation}) into Eq. (\ref{eqenergies}) and considering only massive particles $\delta_i=1$ we obtain three conditions that the parameters of the three particles involved in the decay should fulfill,
\begin{widetext}
\begin{equation}\label{energies2}
\begin{aligned}
E_{1}&=-q_1 A_{t}+\sqrt{-\frac{g_{tt}}{g_{\phi\phi}}}\sqrt{L_1^2+m_1^2 g_{\phi\phi}}, \quad E_{2}=-q_2A_{t}+\frac{m_{0}^2-m_1^2-m_2^2}{2m_1^2}\left(E_1+q_1 A_t\right)-\frac{\epsilon D_0 m_0^2}{2m_1^2}\sqrt{\left(E_1+q_1 A_t\right)^2+m_1^2 g_{tt}},\\
L_2&=\frac{m_0^2-m_1^2-m_2^2}{2m_1^2}L_1 -\frac{\epsilon D_0m_0^2}{2m_1^2}\sqrt{L_1^2+m_1^2g_{\phi\phi}},\quad D_0=\sqrt{\left(1-\left(\frac{m_1+m_2}{m_0}\right)^2\right)\left(1-\left(\frac{m_1-m_2}{m_0}\right)^2\right)}, \quad \epsilon=\pm1.
\end{aligned}
\end{equation}
\end{widetext}
where $E_i=m_i \mathcal{E}_i$, $L_i=m_i l_i$ and $q_i=m_i e_i$ are the energy, angular momentum and charge, respectively. 
The energy $E_0$, angular momentum $L_0$ and charge $q_0$ are given by the conservation laws,
\begin{equation}\label{conlaws}
E_0=E_1+E_2, \quad L_0=L_1+L_2,\quad q_0=q_1+q_2.
\end{equation}
In Eq. (\ref{energies2}) there are six free parameters: the charges $q_1$ and $q_2$, the angular momentum $L_1$, and the masses $m_0$, $m_1$ and  $m_2$. Additionally, from  $D_0$ the masses are restricted to,
\begin{equation}
m_0 \geq m_1+m_2.
\end{equation}

On the other hand, using Eq. (\ref{eff1}) and (\ref{energies2}) we can obtain the general expression for the efficiency of the extraction process, $\eta$, as
\begin{widetext}
\begin{equation}\label{eff3}
\eta = \frac{2m_1^2 E_1}{2m_1^2\left(q_1+q_2\right)A_t-\left(m_0^2+m_1^2-m_2^2\right)\left(E_1+q_1A_t\right)+\epsilon D_0 m_0^2\sqrt{\left(E_1+q_1A_t\right)^2+m_1^2 g_{tt}}},
\end{equation}
\end{widetext}
The advantage of Eq.\ (\ref{energies2}) lies in the explicit dependence on the parameters of the particles involved in the extraction process and the break up point $r_*$, and then we can determine the efficiency at any break up point using Eq. (\ref{eff3}), for an arbitrary static spherically symmetric charged BH.  In contrast to Eq. (\ref{energies}) that is valid for massive and massless particles, Eq. (\ref{eff3}) is only valid for massive particles. Moreover, from Eq. (\ref{eff3}) the generalized ergosphere can be determined from the condition $\eta=0$.    
The  extracted energy comes from the electric interaction between the charged test particle and the charged BH, such that  if the BH is non-charged, $(A_t=0)$, the efficiency becomes negative,  therefore the energy extraction is not possible; the same as with chargeless test particles $q_i=0$. Finally, an interesting limiting case occurs when the break-up point is located at the event horizon $r_*=r_h$; for a static spherically symmetric BH this imply that $g_{tt}=0$, then  Eq. (\ref{eff3}) yields,
\begin{equation}\label{etamax}
\eta_{r_h} =-\frac{q_1}{q_0},
\end{equation}
it is worth to mention that the efficiency in Eq. (\ref{etamax}) is valid for any static spherically symmetric charged BH, and it occurs if a test particle with negative energy reaches the event horizon. Moreover it is independent of the rest of the parameters of the particles and of the mass and charge of the BH. Additional features of the energies in Eq. (\ref{energies2}) and the efficiency in Eq. (\ref{eff3})  are discussed in the next Section for the Reissner-Nordstr\"om BH.

\subsection{The twofold efficiency}\label{twofold}

From Eq. (\ref{energies2}) in case $L_1 \ne 0$, 
associated to $\epsilon= \pm 1$, there are two different values of $L_2$ and $E_2$, therefore there are as well  two possible efficiencies $\eta_1$, $\eta_2$, for the same set of free parameters ($m_0$, $m_1$, $m_2$, $q_1$, $q_2$, $L_1$) at a given breaking point $r_*$.

In contrast, in the case $L_1 =0$  the process reaches its  maximum efficiency for a given break up point $r_*$. In the limit that the break up point approaches the event horizon $r_*\to r_h$ the efficiencies converge to the same limit given by Eq.\ \ref{etamax}, a value that is independent of $L_i$, $m_i$ and the BH parameters.
%%%%%%%%%%%%%%%%%%%%%%%%%%%%%%%%%%%%%%%%%%%%%%%%%%
\subsection{Efficiency with zero angular momentum $L_1=0$}
As we discussed above, if $L_1=0$  ($l_1=0$) the generalized ergosphere reaches its maximum radius,  the efficiency is unique, and Eqs. (\ref{energies2}) lead to,
\begin{equation}\label{energies3}
\begin{aligned}
&E_1 = -q_1A_t+m_1\sqrt{-gtt},\quad |L_{0}| =|L_{2}|=  \frac{\sqrt{g_{\phi\phi}} m_0^2}{2m_1}D_0,\\
 &\qquad E_2= - q_2 + \frac{\left(m_0^2-m_1^2-m_2^2\right)}{2m_1}\sqrt{-gtt}
\end{aligned}
\end{equation}
where $E_0$ and $L_0$ are given by Eq. (\ref{conlaws}), and the efficiency of Eq. (\ref{eff3}) is,
\begin{equation}\label{eff2}
\eta =\frac{2m_1E_1}{2m_1(q_1+q_2)A_t-\left(m^2_0+m^2_1-m^2_2\right)\sqrt{-g_{tt}}}.
\end{equation}

%%%%%%%%%%%%%%%%%%%%%%%%%%%%%%%%%%%%%%%%%%%%%%%%%%
\subsubsection{Efficiency with $D_0=0$}
Another  case where the  efficiency has a unique value is when $D_0=0$, this imply that $m_0=m_1+m_2$, and  Eqs. (\ref{energies2}) lead to,
\begin{equation}
\begin{aligned}
&E_{1}=-q_1 A_{t}+\sqrt{-\frac{g_{tt}}{g_{\phi\phi}}}\sqrt{L_1^2+m_1^2 g_{\phi\phi}},\\
E_{2}&=-q_2A_t+\frac{m_2}{m_1}\left(E_1+q_1 A_t\right),\quad L_2=\frac{m_2}{m_1}L_1,
\end{aligned}
\end{equation}
where $E_0$ and $L_0$ are given by Eq. (\ref{conlaws}); while the efficiency, Eq. (\ref{eff3}) takes the form,
\begin{equation}\label{etad0}
\eta=\frac{m_1 E_1}{m_1\left(q_1+q_2\right)A_t-m_0\left(E_1+q_1A_t\right)}.
\end{equation}
Note that in this case the value of $L_2$ is independent of the break-up point $r_*$ and is directly proportional to the angular momentum $L_1$. This condition produces a lower efficiency than for the case $L_1=0$.
%%%%%%%%%%%%%%%%%%%%%%%%%%%%%%%%%%%%%%%%%%%%%%%%%%%%%%%%%%%%%%%%
\subsection{How the efficiency depends on the masses}\label{masses}
As we mentioned above the masses of the particles are restricted to the condition $m_0 \geq m_1+m_2$.  If $m_0 = (m_1+m_2)$ the efficiency is given by Eq.\ (\ref{etad0}); in this case  the efficiency $\eta$ may be larger or smaller than the efficiency for $m_0 > (m_1+m_2)$. To understand this effect we consider the efficiency for fixed $r_{*}$, $m_0$ and $m_1$; then $m_2$ is in the range $0\leq m_2\leq m_0-m_1$. Deriving  Eq.\ (\ref{eff3}) with respect to $m_2$ we obtain
\begin{equation}\label{deta}
\begin{aligned}
&\frac{d\eta}{dm_2}=\frac{2\epsilon m_2 E_1\left(L_1+L_2\right) D_0\sqrt{-g_{tt}g_{\phi\phi}}}{m_0^2\left(\left(q_1+q_2\right)A_t\sqrt{g_{\phi\phi}}+\epsilon \sqrt{-g_{tt}} C_0\right)^2},\\
&C_0=L_2\left(1+\frac{m_1^2-m_2^2}{m_0^2}\right)-L_1\left(1-\frac{m_1^2-m_2^2}{m_0^2}\right),
\end{aligned}
\end{equation}
with the break-up point $r_*$ inside the generalized ergosphere it is guaranteed  that $E_1<0$ and the sign of the derivative in Eq.\ (\ref{deta}) is determined by the factor $-\epsilon\left(L_1+L_2\right)$. Therefore there are three possible cases if $m_0=m_1+m_2$. The first case is when $L_1=0$ that leads to $d \eta/dm_2>0$, this means that the efficiency $\eta$ is a monotonically increasing function of $m_2$, reaching the maximum  when $m_0=m_1+m_2$ [this is illustrated in Fig.\ \ref{LAFIG5} for the RN BH]. 

For $L_1 \ne 0$ we can choose $\epsilon = \pm 1$. For $\epsilon=-1$,  $d\eta/dm_2>0$, and we conclude again that the efficiency is an increasing function with respect to $m_2$. This means that the maximum efficiency for a given angular momentum $L_1$ and break up point $r_*$, is reached when $m_2$ takes its highest possible value that occurs when $m_0=m_1+m_2$. 
The last case occurs when the angular momentum $L_1 \ne 0$ and $\epsilon=+1$, the sign that the derivative takes depends on the angular momentum $L_1$ and the break-up point $r_*$;  the sign of the derivative is negative if, 
\begin{equation}\label{ep1}
g_{\phi\phi} \le \frac{4L_1^2}{m_0^2D_0^2},
\end{equation}
in this case the efficiency as a function of $m_2$, Eq.\ (\ref{eff3}), is a decreasing function, hence the minimum efficiency corresponds to $m_0= (m_1+m_2)$, and therefore the maximum efficiency occurs when $m_2$ approaches zero. Nevertheless, when Eq.\ (\ref{ep1}) satisfies the equality a maximum efficiency for $L_1 \ne 0$ is reached with $m_2$ given by,
\begin{equation}\label{m2condition}
m_2=\sqrt{m_0^2+m_1^2-\frac{2m_0\sqrt{g_{\phi\phi}\left(L_1^2+m_1^2 g_{\phi\phi}\right)}}{g_{\phi\phi}}}.
\end{equation}
Note that $m_2$, in Eq.\ (\ref{m2condition}), takes real values only if the angular momentum fulfils the condition,
\begin{equation}\label{lcondition}
0<L_1^2\leq\frac{\left(m_0^2-m_1^2\right)^2g_{\phi\phi}}{4m_0^2}.
\end{equation}
On the other hand, the efficiency will be a strictly decreasing function as we mentioned above.\\
The described features of the energy extraction 
become clear in the next section where the process for the RN BH is analyzed.
%%%%%%%%%%%%%%%%%%%%%%%%%%%%%%%%%%%%%%%%%%%%%%%%%%%%%%%%%%%%%%%%%%%%%%%%

\section{Energy extraction of the Reissner-Nordstr\"om BH}\label{RNextr}
The Reissner-Nordstr\"om (RN) metric \cite{Reissner1916,Nordstrom1918} describes a static spherically symmetric charged BH. The metric is given by (\ref{smetric}) with metric functions
\begin{equation}\label{RNmetric}
\begin{aligned}
& g_{tt}=-\frac{\Delta}{r^2},\quad g_{rr}=\frac{r^2}{\Delta},\quad g_{\theta\theta}=r^2,\\
&g_{\phi\phi}=r^2 \sin^2{\theta}, \quad \Delta=r^2-2M r+Q^2,
\end{aligned}
\end{equation}
where $Q$ and $M$ are  the electric charge and mass of the BH, respectively.
The electric potential is 
\begin{equation}\label{RNpot}
A_{t}=-\frac{Q}{r}.
\end{equation}
The event horizon $r_h$ is given by 
\begin{equation}\label{RNhorizon}
r_h = M + \sqrt{M^2-Q^2}.
\end{equation}

Considering massive test particles  $\delta = 1$, the effective potential $V_{+}$ is,
\begin{equation}\label{RNeffpot}
V_{+}=\frac{\sqrt{\Delta(l^2 +r^2 )}}{r^2}+\frac{ e Q}{r}.
\end{equation}
$V_+$ has two contributions, the first one from the gravitational interaction, that is positive definite; the second one is the electrostatic interaction that can be positive or negative according to the sign of $eQ$. Therefore, negative energies ($V_{+} = \mathcal{E} <  0$) are obtained if $eQ<0$, but also is required that the electrostatic interaction dominate over the gravitational one in the range $r_h < r < r_e$. And $r_e$ is defined from the condition $V_+=0$,
\begin{equation}\label{RNcondition}
r^2\left(\Delta-Q^2 e^2\right)-l^2\Delta=0.
\end{equation}

Deriving respect to the angular momentum $l$, we  find that   $r_{e}$ is maximized if $l=0$. Solving for this condition  Eq. (\ref{RNcondition}) leads to \cite{DenardoRuffini1972},
\begin{equation}\label{rereissner}
r_{e_\pm}=M\pm\sqrt{M^2-Q^2\left(1-e^2\right)},
\end{equation}
 where $r_{e_+}$ denotes the boundary of the region where NES can exist, corresponding to  $l=0$. On the other hand, $r_{e_-}$ is located within the event horizon then it is not relevant for the extraction process. 
 
 For angular momentum $l \neq 0$ the radius of the generalized ergosphere is delimited by the event horizon $r_h$ and $r_{e_+}$ that is given by
 \begin{equation}
\begin{aligned}
r_{e}&=\frac{M}{2}+\frac{\sqrt{\alpha}+\sqrt{\alpha-4\beta}}{2},\quad \beta=\frac{a_1+\alpha}{2}+\frac{a_2}{2\sqrt{\alpha}},\\
\alpha& = M^2-2a_{3}+2\sqrt{a_o}\cos\Bigr[\frac{1}{3}\arccos\left(\frac{b_o}{a_o^{3/2}}\right)\Bigr],\\
b_o&=a_3^3+l^2\left(4\left(M^2-Q^2\right)a_3+2M^2e^2Q^2\right),\\
a_o&=a_3^2-\frac{4l^2\left(M^2-Q^2\right)}{3}, \quad a_1=-\frac{3M^2}{2}+3a_3,\\
a_2&=M\left(3a_3-M^2-l^2\right),\quad a_3=\frac{l^2-Q^2\left(e^2-1\right)}{3}.
\end{aligned}
\end{equation}
When $l=0$, the solution $r_{e_+}$ of Eq. (\ref{rereissner}) is recovered. 

\begin{figure}[ht]
\includegraphics[width=8cm,height=5cm]{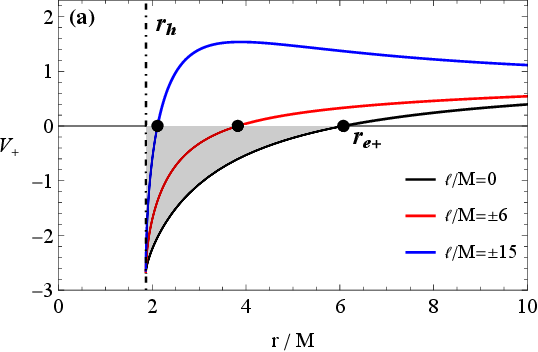}
\quad
\includegraphics[width=8cm,height=5cm]{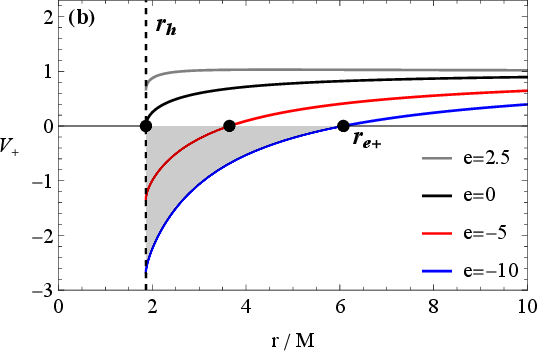}
\caption{\label{FIG1} For the RN BH are illustrated the effective potentials with negative energy states (shaded region) the condition $eQ<0$ is fulfilled with $Q=0.5M$ and $e=-10$ . The region where NES can exists lies between the event horizon $r_h$ and $r_{e_+}$, denoted with the black dots.
(a) varying $l$ as shown $l=0, \pm 6, \pm 15$   (b) for different values of charge mass ratio $e$.  The outer radius of the ergosphere $r_{e_+}$ is larger for larger (negative) values of the charge mass ratio $e$.} 
\end{figure}

In Fig. \ref{FIG1}(a) is shown the effective potential for the particle 1, Eq. (\ref{RNpot}), with NES for different values of the angular momentum; the region where $V_+ < 0$ is $r_h < r < r_{e_+}$. If the angular momentum increases then $r_e$ decreases, and consequently the generalized ergosphere decreases.  In the limit $|l|\to\infty$ the value of $r_{e}$ tends to $r_h$, and  the generalized ergosphere eventually disappears. This happens when the gravitational term in the effective potential Eq. (\ref{RNeffpot}) dominates over the electrostatic interaction, and  the effective potential becomes positive. Note that the sign of the angular momentum is irrelevant in the effective potential $V_+$. In Fig.\ \ref{FIG1}(b) are shown the effective potentials for different values of the charge mass ratio $e_1$; NES exist only if the condition $eQ<0$ is satisfied, otherwise the energy extraction is not possible.

\begin{figure}[htb]
\includegraphics[width=6.5cm,height=6.5cm]{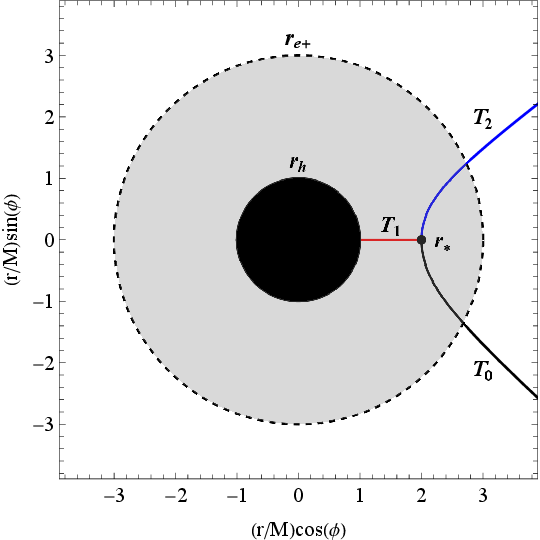}
\caption{\label{FIG2} Illustration of the energy extraction from an extreme RN BH, $Q=M$. The trajectory $T_0$ corresponds to the ingoing particle coming from infinity;  $T_1$ is the trajectory of the particle with negative energy that falls into the black hole and $T_2$ is the trajectory of the outgoing particle that escapes with more energy than the ingoing particle.  The break-up point is at $r_*=2M$; the shadow region represents the generalized ergosphere its outer radius being $r_{e_+}=3M$. The event horizon is $r_h=M$. The parameters of the test particles are given in Table \ref{tabla1}.}
\end{figure} 

We illustrate other features of the energy extraction with the extreme RN BH, that is the RN metric Eq. (\ref{RNmetric}) with $M=Q$; we take the parameters for the extraction process considered in \cite{DenardoRuffini1972}, that  satisfy
Eq. (\ref{energies}).  
When the trajectory $T_0$ coming from infinity and entering the ergosphere reaches a turning point $r_*=2M$, the test particle  splits into two particles that follow the trajectories $T_1$ and $T_2$; $T_1$ goes inside the BH and $T_2$  goes back to infinity as is shown in Fig. \ref{FIG2}. These parameters are listed in Table \ref{tabla1}.

\begin{figure}[ht]
\includegraphics[width=8.5cm,height=5.3cm]{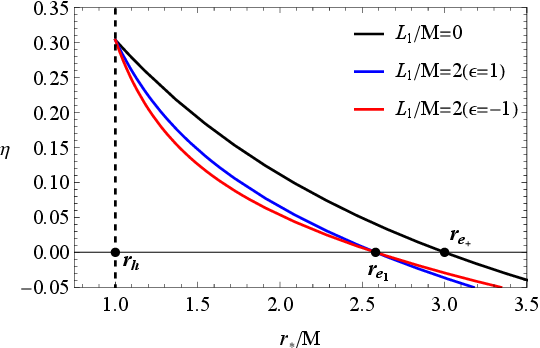}
\caption{\label{FIG3} Efficiency of the extraction process from RN BH via electrostatic interaction as a function of the break-up point $r_*$  for fixed parameters  $Q=M$, $m_0=2.197$, $m_1=1$, $m_2=1$, $q_0=6.586$, $q_1=-2$ and $q_2=8.586$, the charge mass ratio for each particle is  $e_i=q_i/m_i$. The radii of the event horizon and of the generalized ergosphere are $r_h=M$, and  $r_{e+}=3M$ and $r_{e_1}=2.581M$ for $L_1/M=0$ and $L_1/M=2$, respectively.}
\end{figure}

\noi The case analyzed in \cite{DenardoRuffini1972} for a RN BH corresponds to $L_1=0$. For the parameters $q_i$ and $m_i$ in Table \ref{tabla1}, the efficiency is shown as a function of the break-up point. 
It is clear that energy extraction is only possible when the break-up point is located inside the generalized ergosphere, otherwise the efficiency is negative; this means that if the break-up point is outside the ergosphere the outgoing particle carries less energy than the incident particle and there is not energy extraction. 
In Fig. \ref{FIG3}, it is shown that the maximum efficiency is reached when the break-up point $r_*$ is the closest to the event horizon  $r_h$; in this case  the efficiency is given by Eq. (\ref{etamax}). It is worth to mention that the efficiency for $L_1=0$ is larger than for the case $L_1 \ne 0$,  for a given break-up point in the generalized ergosphere. The extraction process described in Table \ref{tabla1} has an efficiency of $\eta=0.1111$.

%%%%%%%%%%%%%%%%%%%%%%%%%%%%%%%%%%%%%%%%%%%%%%%%%%%%%%%%%%%%%
\begin{table}[t]
\centering
\caption{The  parameters considered to illustrate the energy extraction from an extreme RN BH, with $Q=M$. The break-up point is at $r_*=2M$ and the outer radius of the ergosphere is $r_{e_+}=3M$. The event horizon is $r_h=M$. This set of  parameters satisfy Eq. (\ref{energies}).} \label{tabla1}
\begin{ruledtabular}
\begin{tabular}{c c c c c c}
$i$ & $m_{i}$      & $q_{i}$  &$E_{i}$      &$L_{i}/M$&\\ \hline
0       & 2.197         & 6.586      & 4.5             & 2\\
1       & 1               & -2           & -0.5           & 0\\
2      & 1               & 8.586       & 5                & 2          
\end{tabular}
\end{ruledtabular}
\end{table}

In Fig. \ref{FIG3}  we illustrate the twofold efficiency with the parameters in Table \ref{tabla1} for $L_1 =0$, and $L_1=2$ with $\epsilon = \pm 1$; the radii $r_{e+}$ and $r_{e_1}$ delimit the generalized ergosphere. We consider other two scenarios for the parameters  in Table \ref{tabla2} and \ref{tabla3} with  $L_1 =0, 1, 2$.
 
Additionally, notice that a larger efficiency does not imply that the energy of the outgoing particle is greater, in fact, from Table \ref{tabla2} and \ref{tabla3}, for $\epsilon=+1$ the energies $E_0$ and $E_2$ are smaller than for $\epsilon=+1$ but the efficiency is larger.\\

\begin{figure}[ht]
\includegraphics[width=6.8cm,height=6.8cm]{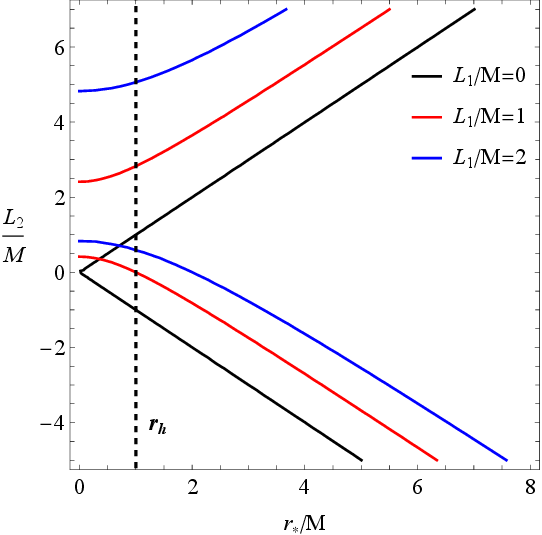}
\caption{\label{FIG4} The angular momentum of the outgoing particle, $L_2$ as a function of $r_*$, in the energy extraction from an extreme RN BH. For $L_1=0$ $L_2$ depends linearly on  $r_*$. On the other hand  if $L_1 \ne 0$  $L_2$ has two values, and two different scenarios arise with different efficiencies. The fixed parameters are $Q=M$, $m_0=2.197$, $m_1=1$, $m_2=1$, $q_0=6.586$, $q_1=-2$ and $q_2=8.586$. The dashed line represents the event horizon $r_h=M$.}
\end{figure}

%%%%%%%%%%%%%%%%%%%%%%%%%%%%%%%%%%%%%%%%%%%%%%%%%%%%%%%%%%%%
\begin{table}[htb]
\centering
\caption{Parameters of the energy extraction from the extreme RN BH, $Q=M$. The break-up point is at $r_*=2M$ and the radius of the ergosphere is $r_{e_1}=2.581M$. The event horizon is $r_h=M$. These parameters fulfill Eq. (\ref{energies2})  with $\epsilon=-1$.} \label{tabla2}
\begin{ruledtabular}
\begin{tabular}{c c c c c c}
$i$ & $m_{i}$      & $q_{i}$  &$E_{i}$      &$L_{i}/M$&\\ \hline
0       & 2.197         & 6.586      & 5.49896            & 7.6512\\
1       & 1               & -2           &-0.292893           & 2\\
2      & 1               & 8.586       & 5.79185              & 5.6512         
\end{tabular}
\end{ruledtabular}
\end{table}
\begin{table}[htb]
\centering
\caption{Parameters for the energy extraction from an extreme RN BH, $Q=M$. The break-up point is at $r_*=2M$ and the radius of the ergosphere is $r_{e_1}=2.581M$;  the event horizon is $r_h=M$. These parameters satisfy Eqs. (\ref{energies2})  with $\epsilon=1$.} \label{tabla3}
\begin{ruledtabular}
\begin{tabular}{c c c c c c}
$i$ & $m_{i}$      & $q_{i}$  &$E_{i}$      &$L_{i}/M$&\\ \hline
0       & 2.197         & 6.586      & 4.50011             & 2.00162\\
1       & 1               & -2           &-0.29289           & 2\\
2      & 1               & 8.586       & 4.793                & 0.00162    
\end{tabular}
\end{ruledtabular}
\end{table}
%%%%%%%%%%%%%%%%%%%%%%%%%%%%%%%%%%%%%%%%%%%%%%%%%%%%%%%%%%
\noi It is important to mention that the two possible efficiencies for a given value of $L_1$ correspond to the two values that the angular momentum $L_2$ can take; for RN BH we can understand this dependence examining  Eq. (\ref{energies2}) for $L_2$.  In the RN BH case 
$L_2$ and break up point $r_*$ satisfy a hyperbola equation,  
\begin{equation}
\left(L_2-\frac{m_0^2-m_1^2-m_2^2}{2m_1^2}L_1\right)^2-\frac{m_0^4 D_0^2}{4m_1^2}r_*^2 =\frac{m_0^4D_0^2}{4m_1^4}L_1^2.
\end{equation}
In Fig. \ref{FIG4} it is shown that when $L_1 \ne 0$ there are two values of the  angular momentum $L_2$ for a given break-up point $r_*$; therefore there are two associated values of $E_2$ and $\eta$. On the other hand, when $L_1 =0$, $L_2$ (and therefore $L_0$) depends linearly on  $r_*$; also note that if 
$L_1=0$ although $L_2$ (with $L_1=0$) can take two values, both have the same magnitude, namely, there is no split of efficiency for $L_1=0$. 
%%%%%%%%%%%%%%%%%%%%%%%%%%%%%%%%%%%%%%%%%%%%%%%%%%%%%%%%%%%%%%%%%%
%%%%%%%%%%%%%%%%%%%%%%%%%%%%%%%%%%%%%%%%%%%%%%%%%%%%%%%%%%%%%
%%%%%%%%%%%%%%%%%%%%%%%%%%%%%%%%%%%%%%%%%%%%%%%%%%%%%%%%%%%%%%%%%%
\begin{figure}[ht]
\includegraphics[width=8cm,height=5cm]{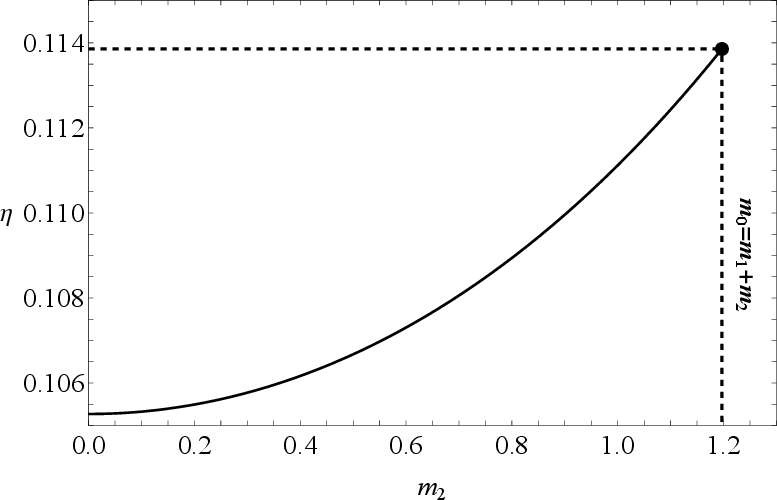}
\caption{\label{LAFIG5} The efficiency as a function of $m_2$ for $L_1=0$ in the extreme RN BH,  $Q=M$. The radius of the event horizon is $r_h=M$. The parameters of the test particles are $m_0=2.197$, $m_1=1$, $q_1=-2$ and $q_2=8.586$. $m_2$ is in the range $0 \le m_2 \le (m_0-m_1)$. The efficiency takes its maximum value when $m_0=m_1+m_2$.}
\end{figure} 
%%%%%%%%%%%%%%%%%%%%%%%%%%%%%%%%%%%%%%%%%%%%%%%%%%%%%%%%%%%%%%
 The dependence of the efficiency  on the masses for RN BH is illustrated in  Figs.\ref{LAFIG5} and  \ref{FIG6}.
 In Fig. \ref{FIG6} is shown the efficiency for a given break-up point $r_*$ and angular momentum $L_1$ for $\epsilon= \pm 1$. In both figures we can observe that when $\epsilon=-1$ the efficiency is an increasing function that reaches its maximum value when $m_0=m_1+m_2$. In the case $\epsilon=+1$, if the angular momentum does not satisfy Eq.\ (\ref{lcondition}) the efficiency is a decreasing function and its minimum  occurs for $m_0=m_1+m_2$, as is shown in Fig.\ \ref{FIG6}a.  If the angular momentum $L_1$ fulfills  Eq. (\ref{lcondition}) the efficiency reaches its maximum value when $m_2$ satisfies Eq.\ (\ref{m2condition}). In the case $m_0=m_1+m_2$, the efficiencies for $\epsilon=\pm 1$ tend to the one given in Eq.\ (\ref{etad0}).
%%%%%%%%%%%%%%%%%%%%%%%%%%%%%%%%%%%%%%%%%%%%%%%%%%%%%%%%%%%%%%%%
\begin{figure}[ht]
\includegraphics[width=8cm,height=5cm]{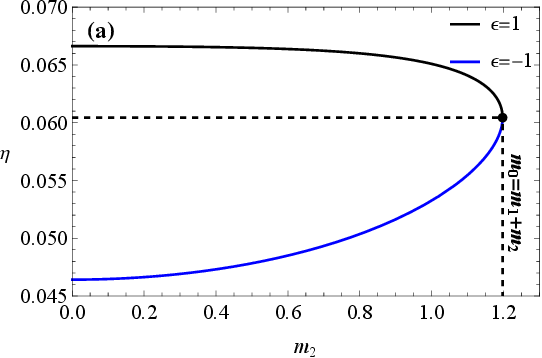}
\quad
\includegraphics[width=8cm,height=5cm]{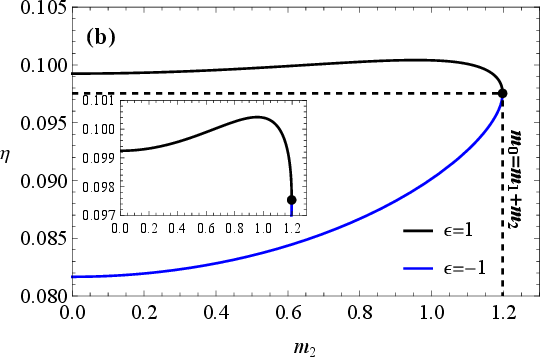}
\caption{\label{FIG6} The efficiency as a function of $m_2$ for the extreme RN BH,  $M=Q$. The radius of the event horizon is $r_h=M$. The fixed parameters of the particles are $m_0=2.197$, $m_1=1$, $q_1=-2$ and $q_2=8.586$. While $m_2$ is in the range  $0 \le m_2 \le m_0-m_1$. (a) For $L_1=2$ the efficiency with $\epsilon=+1(\epsilon=-1)$ is a decreasing (increasing) function and the minimum (maximum)  occurs for $m_0=m_1+m_2$, black (blue) curve. (b) For $L_1=1$, the efficiency for $\epsilon=-1$ is an increasing function (blue line) an its maximum occurs for $m_0=m_1+m_2$, while for $\epsilon=+1$ the maximum  occurs for $m_2=0.956121$ (in the zoom).}
\end{figure} 
%%%%%%%%%%%%%%%%%%%%%%%%%%%%%%%%%%%%%%%%%%%%%%%%%%%%%%%%%%%%%%%%%%%%%%%
\section{Reissner Nordstr\"om de Sitter Black Hole,   $\Lambda \ge 0$}\label{secRNLamnda}

 The spacetime that describes a Reissner Nordstr\"om BH with cosmological constant $\Lambda$ is given by Eq. (\ref{smetric}) with metric coefficients in coordinates ($t$, $r$, $\theta$, $\phi$) are \cite{Cai2002}
 \begin{equation}\label{RNLmetric}
 g_{tt}=-\frac{\Delta}{r^2},\quad  g_{rr}=\frac{r^2}{\Delta},\quad g_{\theta\theta}=r^2, \quad g_{\phi\phi}=r^2\sin^2\theta,
 \end{equation}
 where
 \begin{equation}
\Delta = r^2-2M r+Q^2- \frac{\Lambda}{3}r^4,
 \end{equation}
and $M$ and $Q$ are the mass and charge of the BH, and $\Lambda$ is the cosmological constant. The RN-de Sitter (RNdS) and RN-anti-de Sitter (RN-AdS) correspond to positive or negative cosmological constant, respectively, and is non-asymptotically flat. The event horizons are given by the solution of $g_{rr}^{-1}=0$, equivalently $\Delta=0$, or $g_{tt}=0$. In contrast with the horizons of RN BH,  Eq. (\ref{RNhorizon}), the RN-$\Lambda$ BH has the possibility of  two or three horizons depending on the sign of $\Lambda$. 

\subsection{Horizons of the RNdS BH} 
\begin{figure}[ht]
\includegraphics[width=8cm,height=5cm]{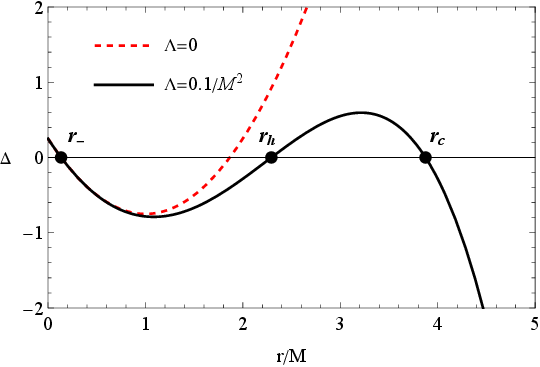}
\caption{\label{FIG7}  $\Delta(r)$  for RN BH (dashed line) and RNdS (solid line) BH. The parameters have been fixed as $Q=0.5M$ and $\Lambda = 0.1/M^2$. The dots correspond to the location of the cosmological horizon $r_c=3.8749M$, the event horizon $r_h=2.2926M$ and the inner horizon $r_-=0.1339M$.}
\end{figure}

From $\Delta =0$ are determined the cosmological, outer and inner horizons,  $r_c$, $r_h$,  $r_-$, respectively,
\begin{equation}\label{deltarnds}
 \Delta = r^2-2M r+Q^2- \frac{\Lambda}{3}r^4=0,
 \end{equation}
 the roots given by
 \begin{equation}\label{horizons}
 \begin{aligned}
 & r_c=\alpha +\sqrt{\alpha^2-\beta_+}, \qquad\quad r_h=\alpha -\sqrt{\alpha^2-\beta_+},\\
 & r_- =-\alpha +\sqrt{\alpha^2-\beta_-}, \qquad \beta_{\pm} = 2\alpha^2-3A_1\pm\frac{3M}{2\alpha\Lambda},\\
&A_1=\frac{1}{2\Lambda},\qquad A_2=\frac{1-4Q^2 \Lambda}{16\Lambda^2}, \qquad A_3=-\left(\frac{3M}{4\Lambda}\right)^2,\\
&\alpha = \sqrt{A_1+2\sqrt{A_2}\cos\Bigr[\frac{1}{3}\arccos\left(\frac{A_1\left(3A_2-A_1^2\right)-A_3}{2A_2^{3/2}}\right)\Bigr]}.\\
 \end{aligned}
 \end{equation}
This roots \footnote{The fourth root is negative, $r_{--}=-\alpha -\sqrt{\alpha^2-\beta_-}$.} satisfy the relation $0<r_-<r_h<r_c$. Fig. \ref{FIG7} exhibits the three horizons for RNdS BH for fixed $M$, $Q$ and positive $\Lambda$, showing also the RN BH case with two horizons,  Eq. (\ref{RNhorizon}). \\
%%%%%%%%%%%%%%%%%%%%%%%%%%%%%%%%%%%%%%%%%%%%%%%%%%%%%%%%%%%%%%%%%%%%%%%
%%%%%%%%%%%%%%%%%%%%%%%%%%%%%%%%%%%%%%%%%%%%%%%%%%%%%%%%%%
\subsubsection{Degeneracy of the horizons}
According to Eq.(\ref{RNhorizon}), the RN BH has two horizons when $Q \leq M$.  If $M=Q$, there is a degenerate horizon $r_h=r_{-}$ and Eq. (\ref{RNmetric}) describes an extreme BH. When $M<Q$ there are no horizons, and Eq. (\ref{RNmetric}) exhibits a naked singularity. However, for RNdS BH the introduction of an additional parameter, $\Lambda>0$, changes this situation and  the relative magnitudes of $M$, $Q$ and $\Lambda$ define the number of horizons.
This deserves a deeper analysis that we carry out in what follows. We  distinguish three cases of degeneracy following the notation in \cite{Sharmanthie2013}.
\subsubsection{Case I: Degeneration of the event horizon and cosmological horizon  $r_c=r_h$}
The first case of degeneracy is the charged Nariai BH that occurs when the event horizon and cosmological horizon coincide $r_h=r_c$, in this case the three horizons of Eq. (\ref{horizons}) take the explicit form,
\begin{equation}
\begin{aligned}
&r_h=r_c =\sqrt{A_1+2\sqrt{A_2}}, \quad r_{-} =-r_h+2\sqrt{A_1-\sqrt{A_2}},\\
&A_1=\frac{1}{2\Lambda},\quad A_2=\frac{1-4Q^2\Lambda}{16\Lambda^2}, \quad M=\frac{4\Lambda r_h\left(A_1-\sqrt{A_2}\right)}{3}.
\end{aligned}
\end{equation}
Taking the mass of the BH $M$ as a free parameter it is possible to establish  the range for the cosmological constant $\Lambda$ and charge $Q^2$ for RNdS BH,
\begin{equation}\label{lambda1}
0\leq \Lambda  \leq \frac{6\left(M+\sqrt{9M^2-8Q^2}\right)}{\left(3M+\sqrt{9M^2-8Q^2}\right)^3}, \quad  0\leq Q^2 \leq \frac{9}{8} M^2.
\end{equation} 

\subsubsection{Case II: Extreme BH: Coalesced event and inner horizons $r_h=r_-$}
The second case of degeneracy occurs when the event and inner horizon coincide $r_h=r_-$; this represents an extreme BH surrounded by a cosmological horizon, and the explicit expressions of the horizons are,
\begin{equation}\label{RNextremeL}
\begin{aligned}
& r_c = -r_h+2\sqrt{A_1+\sqrt{A_2}}, \quad r_h=r_-=\sqrt{A_1-2\sqrt{A_2}},\\
&M = \frac{4}{3}\Lambda r_h\left(A_1+\sqrt{A_2}\right), \quad A_1=\frac{1}{2\Lambda}, \quad A_2 =\frac{1-4Q^2\Lambda}{16\Lambda^2}.
\end{aligned}
\end{equation}
The expansion in series around of $\Lambda\ll1$ in the mass condition of Eq. (\ref{RNextremeL}), leads to
\begin{equation}\label{conditionL}
\frac{M}{Q}=1-\frac{Q^2}{3}\left(\frac{\Lambda}{2}\right)-\frac{Q^4}{2}\left(\frac{\Lambda}{2}\right)^2-\frac{3Q^6}{2}\left(\frac{\Lambda}{2}\right)^3+\mathcal{O}\left(\left(\frac{\Lambda}{2}\right)^4\right),
\end{equation}
where is evident that for $\Lambda=0$, the condition of the extreme RN BH $M=Q$ is recovered; however, for $\Lambda \ne 0$,  Eq. (\ref{conditionL}) suggests that the charge $Q$ can exceed the mass $M$, $Q>M$,  for and extreme RNdS BH.

\subsubsection{Case III: Three degenerate horizons $r_c = r_h = r_-$}
The third case occurs when the three horizons coincide $r_c = r_h = r_-$; in this scenario the horizons are given by,
\begin{equation}\label{3Dhorizon}
r_c=r_h=r_-=\sqrt{\frac{1}{2\Lambda}}, 
\end{equation}
when the three horizons degenerate into one, the value of the single horizon can be written down in terms of a single parameter $M$, $Q$ or $\Lambda$, using the relations
\begin{equation}
Q=\frac{1}{2\sqrt{\Lambda}}, \quad M=\sqrt{\frac{1}{2\Lambda}}.
\end{equation}
%%%%%%%%%%%%%%%%%%%%%%%%%%%%%%%%%%%%%%%%%%%%%%%%%%%%%%%%%%%%%%%%%%%%%%
\subsection{Generalized ergosphere of the RNdS BH for zero angular momentum}\label{gergo}
To obtain the generalized ergosphere is necessary to determine the regions where the effective potential in Eq. (\ref{potefectivo}) becomes negative;  the condition to solve is  Eq. (\ref{regcondition}). As we discussed in Sec. \ref{secergo} the maximum radius of the generalized ergosphere is reached when the angular momentum is zero, then imposing $l=0$ and using Eqs. (\ref{regcondition}) and (\ref{RNLmetric}) it is obtained
\begin{equation}
r^2-2Mr +\mathcal{Q}-\frac{\Lambda}{3}r^4=0, \quad \mathcal{Q}=Q^2\left(1-e^2\right),
\end{equation}
%%%%%%%%%%%%%%%%%%%%%%%%%%%%%%%%%%%%%%%%%%%%%%%%%%%%%%%%%%%%%%%
\begin{figure*}[ht]
\includegraphics[width=8cm,height=5cm]{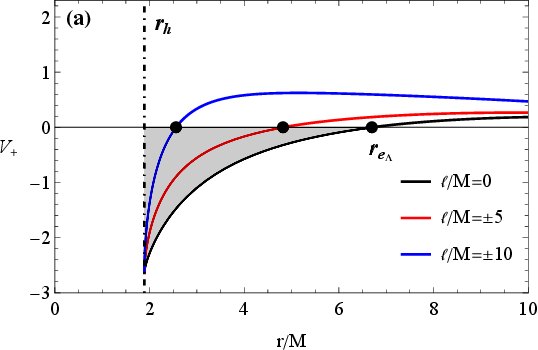}
\includegraphics[width=8cm,height=5cm]{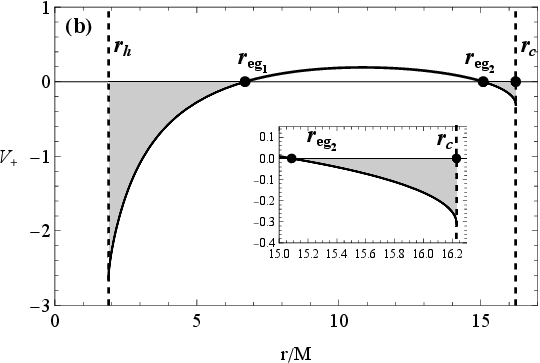}
\caption{\label{FIG8} The effective potential for RNdS BH; the BH parameters fixed as $Q=0.5M$, $\Lambda=0.01/M^2$, and for the test particle $m=1$, $q=-10$. The grey area is the ergosphere. (a) The vertical dot-dashed line represents the event horizon $r_h$ of RNdS BH; the dots are the radii of the generalized ergosphere for RN BH and RNdS BH labelled as $r_e$ and $r_{e\Lambda}$, respectively. The introduction of $\Lambda$ increases the outer radius of the generalized ergosphere; the largest  ergosphere is for $L=0$ (b) $\Lambda$ produces two regions where negative energy states are allowed; one  first generalized ergosphere in the spherical shell $r_h<r<r_{eg_1}$ and a second one, a cosmological ergosphere, delimited by $r_{eg_2}<r<r_{c}$.}
\end{figure*}
%%%%%%%%%%%%%%%%%%%%%%%%%%%%%%%%%%%%%%%%%%%%%%%%%%%%%%%%%%%%%%%%%%%%%%%%
this condition has the same form as $\Delta=0$ that give us the horizon of the RNdS BH with the change $Q^2\to\mathcal{Q}$. Then the  horizons given by Eqs. (\ref{horizons}) are valid for the generalized ergosphere  changing $Q^2 \to \mathcal{Q}$. Since the radius of the generalized ergosphere must be limited by the event horizon $r_h$ and the cosmological horizon $r_c$, the only physical solutions is,
\begin{equation}\label{geRNL}
\begin{aligned}
 &r_{eg_1}=\alpha -\sqrt{\alpha^2-\beta_+},\qquad\quad  r_{eg_2}=\alpha +\sqrt{\alpha^2-\beta_+},\\
 &\beta_{\pm} = 2\alpha^2-3A_1\pm\frac{3M}{2\alpha\Lambda}, \qquad \mathcal{Q}=Q^2\left(1-e^2\right),\\
&A_1=\frac{1}{2\Lambda},\qquad A_2=\frac{1-4\mathcal{Q} \Lambda}{16\Lambda^2}, \qquad A_3=-\left(\frac{3M}{4\Lambda}\right)^2,\\
&\alpha = \sqrt{A_1+2\sqrt{A_2}\cos\Bigr[\frac{1}{3}\arccos\left(\frac{A_1\left(3A_2-A_1^2\right)-A_3}{2A_2^{3/2}}\right)\Bigr]}.\\
\end{aligned}
\end{equation}
Fig. \ref{FIG8} displays the effective potential for RNdS BH; Fig. \ref{FIG8}(a) shows that the radius of the ergosphere $r_{eg_1}$ of the RNdS BH increases in presence of the cosmological constant in relation with the generalized ergosphere for RN BH, $r_{e+}$. Moreover it is shown that the largest radius of the ergosphere is reached when the angular momentum is $l=0$. In contrast with the Kerr BH or the RN BH where there is only one ergoregion, the existence of  two radius in Eq. (\ref{geRNL}) implies two ergoregions with NES as shown in Fig. \ref{FIG8}(b); the first of them is the generalized ergosphere of BH  in  $r_h<r<r_{eg_1}$. The second one is due to the presence of the cosmological constant $\Lambda>0$ and is  $r_{eg_2} < r < r_{c}$, we will refer to this region as the cosmological esgosphere. The existence of  two ergospheres is discussed in \cite{Sourav2018} for the  Kerr-de Sitter spacetime with chargeless particles.\\

\begin{figure*}[htpb]
\includegraphics[width=8cm,height=5cm]{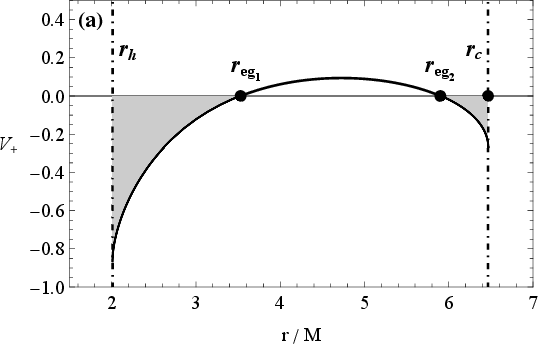}
\quad
\includegraphics[width=5cm,height=5cm]{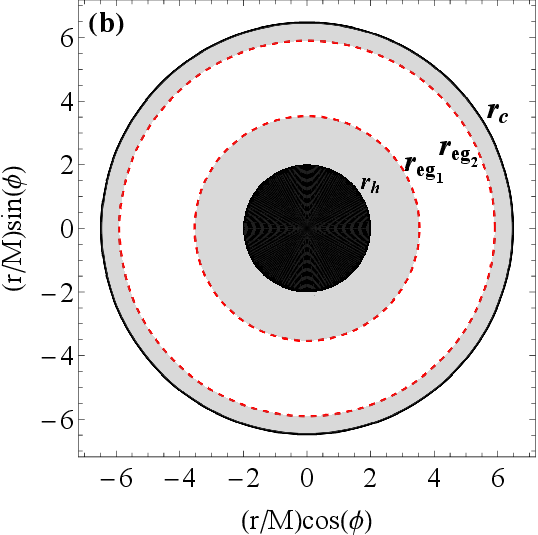}\\
\includegraphics[width=8cm,height=5cm]{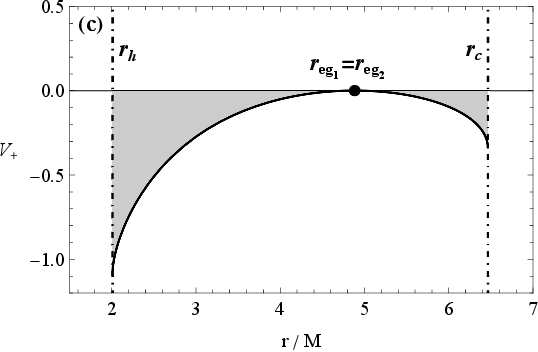}
\quad
\includegraphics[width=5cm,height=5cm]{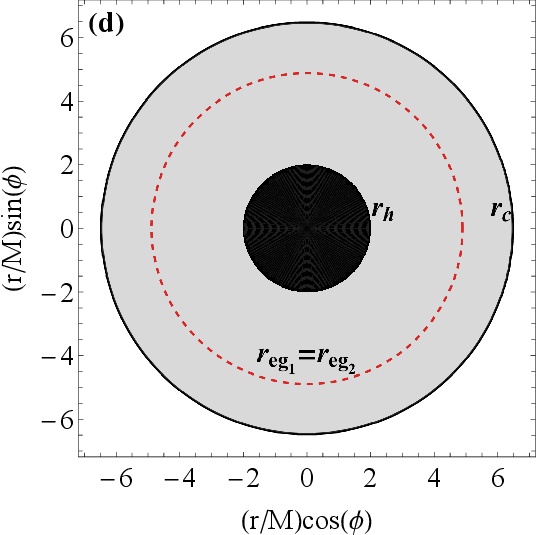}\\
\includegraphics[width=8cm,height=5cm]{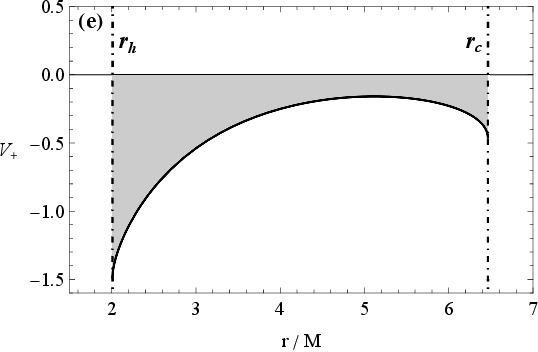}
\quad
\includegraphics[width=5cm,height=5cm]{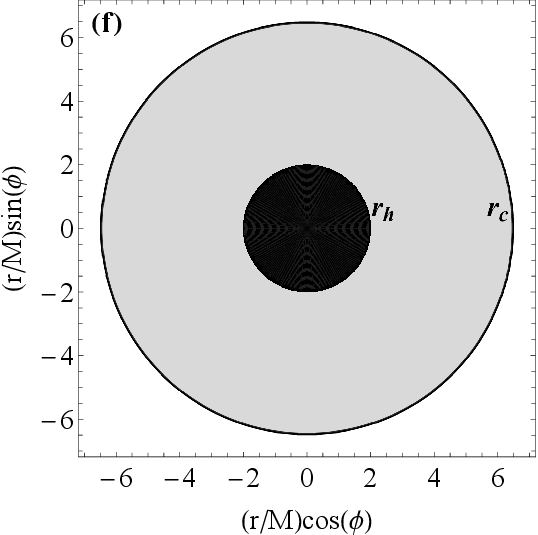}
\caption{\label{FIG9}. The effective potential for RNdS BH in the cases: $e < e_{crit}$, $e = e_{crit}$ and $e > e_{crit}$ are shown. The parameters are fixed as $Q=0.5M$, $\Lambda=0.05/M^2$ and $l=0$; $e_{crit}=-4.4056$, the dots represent the event horizon and cosmological horizon,  $r_h=2.0113M$ and $r_c=6.4651M$, respectively. (a) The effective potential for $e=-3.5$, the black dots represent the radii of the generalized ergosphere where $r_{eg_1}=3.5297M$ and $r_{eg_2}=5.9005M$, (b) the generalized ergospheres $e=-3.5$ in the equatorial plane. (c) The effective potential for $e=e_{crit}=-4.4056$, the black dot represents the outer radius of the ergosphere that is degenerate in $r_{crit}=4.88445$; (d) the generalized ergospheres for $e=e_{crit}=-4.4056$ in the equatorial plane.  (e) The effective potential with $e=-6$, NES are allowed in the range $r_h<r<r_c$ (f) the generalized ergospheres for $e=-6$ in the equatorial plane. The gray area represents the domain of $r$ where NES can exist.}
\end{figure*}

\begin{figure}[H]
\includegraphics[width=8cm,height=5cm]{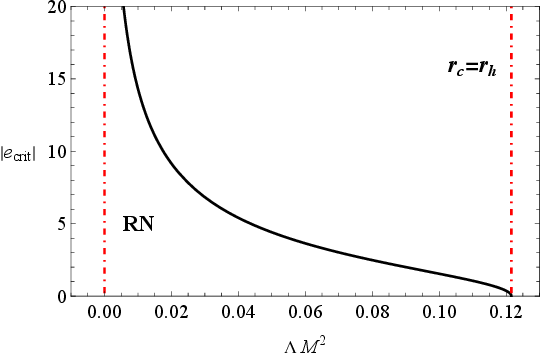}
\caption{\label{FIG10} $e_{crit}$ as a function of $\Lambda$. The left vertical line corresponds to  $|e_{crit}|$ for RN BH ($\Lambda=0$). The right line correspond to $|e_{crit}|$ when the event horizon and cosmological horizon are the same. We fixed $Q=0.5M$.}
\end{figure}

Similar to the case of the RN BH generalized ergosphere, the ergosphere in the RNdS BH depends on the charge mass ratio $e$ of the splitting particle, but an important difference lies in the fact that the cosmological constant generates a second region where NES are allowed, and both regions increase if $e$ increases, and decrease in the opposite case. An interesting scenario occurs when $e$ is large enough and both ergospheres increase until  $r_{eg_1} =r_{eg_2}$ and the effective potential is negative for any  $r$ in the domain $r_h < r < r_c$. The critic value $e_{crit}$ that produces $r_{eg_1}=r_{eg_2}$ is defined by the solution of
\begin{equation}\label{conditioncharge}
\begin{aligned}
&\qquad\quad\mathcal{P}=0, \quad \frac{d\mathcal{P}}{dr}=0,\\
\mathcal{P}&=g_{\phi\phi}\left(\mathcal{E}+eA_t\right)^2+g_{tt}\left(l^2+g_{\phi\phi}\delta\right).
\end{aligned}
\end{equation}
Considering the condition that maximizes the radii of both ergospheres, namely $l=0$, and imposing $E=0$ that implies that $r$ is the radius of the ergosphere,  the solution of Eq. (\ref{conditioncharge}) for the critic charge $e_{crit}$ is,
\begin{equation}
e_{crit} = -\sqrt{1+\frac{r_{crit}(r_{crit}-3M)}{2Q^2}},
\end{equation}
where $r_{crit}$ is the value of $r$ that makes $r_{crit}=r_{eg_1}=r_{eg_2}$ and is given by,
\begin{equation}\label{rcrit}
r_{crit}=\sqrt{\frac{2}{\Lambda}}\cos\left(\frac{1}{3}\arccos\left(-3M \sqrt{\frac{\Lambda}{2}}\right)\right).
\end{equation}
We can distinguish three different scenarios according to the value of $e$ respect to $e_{crit}$, that are shown in Fig. \ref{FIG9}. The first case  occurs when $e < e_{crit}$ and is shown in Fig. \ref{FIG9} (a) where the effective potential exhibits two regions (gray areas) with NES, one of them delimited by $r_h$ and $r_{eg_1}$, and the other one by $r_{eg_2}$ and $r_c$. In Fig. \ref{FIG9}(b) is illustrated the generalized ergosphere for the case $e < e_{crit}$ in the equatorial plane. The second case occurs when the charge mass ratio $e_{crit}$ is reached by the test particle, in this scenario the radii of both ergospheres coalesce at $r_{crit} = r_{eg_1}=r_{eg_2}$. 
%%%%%%%%%%%%%%%%%%%%%%%%%%%%%%%%%%%%%%%%%%%%%%%%%%%%%%%%
%%%%%%%%%%%%%%%%%%%%%%%%%%%%%%%%%%%%%%%%%%%%%%%%%%%%%%%%%%%%%%%%%%%%
Fig. \ref{FIG9}(c) displays the effective potential with allowed NES in the range  $r_h < r < r_c$, except at $r_{crit}$ where the potential vanishes. In Fig. \ref{FIG9}(d) is shown the generalized ergosphere for the case $e=e_{crit}$ in the equatorial plane. 
The third case occurs when the charge mass ratio $e$ exceeds the value of $e_{crit}$,  Figs. \ref{FIG9}(e) and \ref{FIG9}(f) exhibit the effective potential and generalized ergosphere in the equatorial plane when $e > e_{crit}$, in this case the potential is negative in the whole range  $r_h<r<r_c$, in this region the NES are allowed.\\
\noi In Fig.\ \ref{FIG10} is shown how $e_{crit}$ depends on the cosmological constant $\Lambda$,   given by Eq.\ (\ref{lambda1}). When $\Lambda=0$ we recover the case of RN BH (without cosmological horizon). The divergence of $|e_{crit}|$ occurs when the generalized ergosphere covers the whole vicinity of the RNdS BH case. Moreover, for  $\Lambda \ne 0$ the value of $|e_{crit}|$ decreases monotonically until  the inequality in Eq.\ (\ref{lambda1}) is saturated, namely,
\begin{equation}
\Lambda =\frac{6\left(M+\sqrt{9M^2-8Q^2}\right)}{\left(3M+\sqrt{9M^2-8Q^2}\right)^3},
\end{equation}
where $|e_{crit}| = 0$, and the region delimited by the event horizon $r_h$ and the cosmological horizon $r_c$ diminishes.
%%%%%%%%%%%%%%%%%%%%%%%%%%%%%%%%%%%%%%%%%%%%%%%%%%%%%%%%%%
%%%%%%%%%%%%%%%%%%%%%%%%%%%%%%%%%%%%%%%%%%%%%%%%%%%%%%%%%%%%%%%%%%
\noi Finally,  in the case $l \neq 0$ there are as well two regions with NES, while the condition to determine the radius of the generalized ergosphere and the cosmological ergosphere is,
\begin{equation}\label{egcondition2}
r^2\left(\Delta-Q^2 e^2\right)+l^2 \Delta=0,
\end{equation}
where $\Delta$ is given by Eq. (\ref{deltarnds}), and Eq. (\ref{egcondition2}) is a sixth degree polynomial in $r$ that we solve numerically.
%%%%%%%%%%%%%%%%%%%%%%%%%%%%%%%%%%%%%%%%%%%%%%%%%%%%%%%%%%%%%%%%%%%%%
\subsection{Efficiency of the energy extraction in the RNdS BH}

 Considering Eq. (\ref{eff3}) and $g_{tt}$,  Eq. (\ref{RNLmetric}), we analyze the efficiency of the energy extraction from RNdS BH. The first scenario we consider is the case when the ergosphere reaches it maximum for a particle with charge $e_1$ and $L_1=0$. The efficiency in terms of the masses and charges of the particles involved in the process for an arbitrary break-up point $r_*$ can be determined with Eq. (\ref{eff2}); it is only necessary to satisfy the conditions  $m_0 \geq m_1+m_2$ and $e_1 Q<0$.
 %%%%%%%%%%%%%%%%%%%%%%%%%%%%%%%%%%%%%%%%%%%%%%%%%%%%%%%%%%%%%%%%%%%%%%
 \begin{figure}[ht]
\includegraphics[width=8cm,height=5cm]{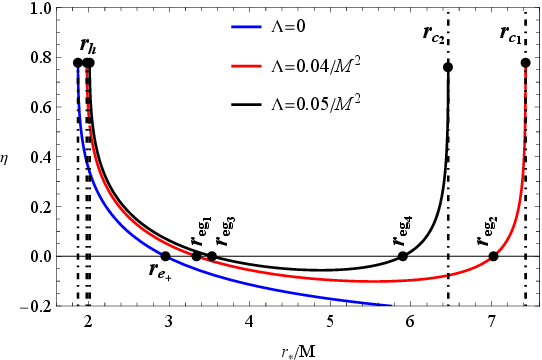}
\caption{\label{LAFIG11} The efficiency of the extraction process as a function of the break up point $r_*$ for RNdS BH for different values of $\Lambda$. The dot-dashed lines represent the event horizons $r_h$ and cosmological horizons $r_c$. The fixed parameters are $Q=0.5M$, $m_0=2.3$, $m_1=1$, $m_2=1.1$, $q_0=4.5$, $q_1=-3.5$ and $q_2=8$. For $\Lambda=0$ the BH presents one event horizon at $r_h=1.86603M$ and the ergosphere radius is $r_{e+}=2.95256M$. For $\Lambda=0.04/M^2$ the BH has one event horizon at $r_h=1.97645M$ and a cosmological horizon at $r_{c_1}=7.425615M$; moreover the radii of the generalized ergosphere are $r_{eg_1}=3.33859M$ and  $r_{eg_2}=7.02733M$. For $\Lambda=0.05/M^2$ the BH has one event horizon at $r_h=2.01131M$ and a cosmological horizon at $r_{c_2}=6.46512M$, moreover the radii of the generalized ergosphere are $r_{eg_3}=3.52977M$ and  $r_{eg_4}=5.90049M$. }
\end{figure} 
%%%%%%%%%%%%%%%%%%%%%%%%%%%%%%%%%%%%%%%%%%%%%%%%%%%%%%%%%%%%%%%
\noi The effect produced by the cosmological constant in the efficiency as a function of $r_*$ is shown in Fig. \ref{LAFIG11}  for different values of $\Lambda$: (i) $\Lambda=0$ (blue curve) corresponds to the efficiency of RN BH, the region where the efficiency $\eta$ is positive lies in the region $r_h<r_*<r_{e+}$, where $r_h=1.86603M$ is the event horizon and $r_{e+}=2.95256M$ is the radius of the generalized ergosphere. (ii) For $\Lambda=0.04/M^2$, we can identify two regions where $\eta>0$,  $r_h<r_*<r_{eg_1}$ and $r_{eg_2}<r_*<r_{c_1}$, where  $r_h=1.97645M$, $r_{c_1}=7.425615M$, $r_{eg_1}=3.33859M$ and  $r_{eg_2}=7.02733M$, that  correspond to the regions where NES are allowed. (iii) For $\Lambda=0.05/M^2$, there are two regions where $\eta>0$,  $r_h<r_*<r_{eg_3}$ and $r_{eg_4}<r_*<r_{c_2}$, where  $r_h=2.01131M$, $r_{c_2}=6.46512M$, $r_{eg_3}=3.52977M$ and  $r_{eg_4}=5.90049M$. Note that $\eta > 0$ grows as $\Lambda$  increases.\\

\begin{figure}[H]
\includegraphics[width=8cm,height=5cm]{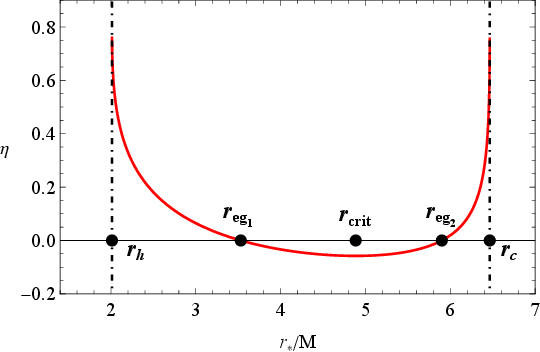}
\caption{\label{FIG12} It is shown the efficiency $\eta$ for the case $e_1 < e_{crit}$, as a function of the break up point $r_*$. the rest of the parameters are fixed: $Q=0.5M$, $\Lambda=0.05/M^2$, $m_0=2.3$, $m_1=1$, $m_2=1.1$, $q_0=4.5$, $q_2=8$, and $q_1=e_1=-3.5$. The critic charge and critic radii are $e_{crit}=-4.4056$ and $r_{crit}=4.8844$, respectively; while the event horizon and cosmological horizon are $r_h=2.0113$ and $r_c=6.4651$.}
\end{figure}

\begin{figure}[ht]
\includegraphics[width=8cm,height=5cm]{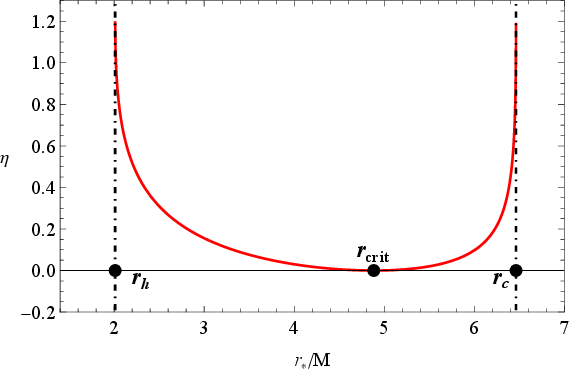}
\caption{\label{FIG13} The efficiency $\eta$ versus the break up point $r_*$ for RNdS in the case $e_1=e_{crit}$. The rest of the parameters fixed as $Q=0.5M$, $\Lambda=0.05/M^2$, $m_0=2.3$, $m_1=1$, $m_2=1.1$,  $q_0=3.5944$, $q_2=8$, and $q_1=e_1=-4.4056$. The critic charge and critic radius are $e_{crit}=-4.4056$ and $r_{crit}=4.8844$, respectively. Both ergospheres join at $r_{crit}$. The event horizon and cosmological horizon are $r_h=2.0113$ and $r_c=6.4651$.}
\end{figure}

Analysing how $\Lambda$ affects the efficiency of the energy extraction, we highlight the existence of two regions where the efficiency is positive for $\Lambda>0$ in constrast with the case $\Lambda=0$, with only one region with $\eta > 0$; consequently,  for $\Lambda > 0$ there are two break up points $r_*$ with the same efficiency. On the other hand, if $r_*$ is outside of the generalized ergosphere then the outgoing particle has less energy than  the ingoing particle and the efficiency becomes negative.  Additionally, $\Lambda > 0$ makes the extraction process more efficient at any break-up point where the NES are allowed in contrast to the case $\Lambda=0$; as $\Lambda$ increases the regions where the efficiency is positive also increase. Finally, it is worth to mention that the maximum efficiency is reached when the break-up point $r_*$ is located at the event horizon or the cosmological horizon; the efficiency at those points is independent of  $\Lambda$ and the rest of the BH parameters $M$ and $Q$, and is given by $\eta_{r_h}=-q_1/q_0$, Eq.(\ref{etamax}).\\
Analogously to Sec. \ref{gergo}, we can distinguish three scenarios for the RNdS BH according to the value of charge mass ratio $e_1$ and $L_1=0$. In Fig. \ref{FIG12}, are illustrated the three possible cases: (i) $e_1 < e_{crit}$ with $\Lambda > 0$ there are two region where the energy extraction is possible. In this scenario $r_{crit}$ is located outside of the ergospheres and represents the break up point where the outgoing particle has less energy than the incident particle.
%%%%%%%%%%%%%%%%%%%%%%%%%%%%%%%%%%%%%%%%%%%%%%%%%%%%%%%%%%%%%%%%%
%%%%%%%%%%%%%%%%%%%%%%%%%%%%%%%%%%%%%%%%%%%%%%%%%%%%%%%%%%%%%%%
\noi(ii)  $e_1 = e_{crit}$, see Fig.\ \ref{FIG13}, in this scenario both ergospheres join at $r_{crit}$ and the entire region delimited by the event horizon $r_h$ and the cosmological horizon $r_c$ is viable for the extraction process, except the point $r_{crit}$ where the efficiency vanishes.
%%%%%%%%%%%%%%%%%%%%%%%%%%%%%%%%%%%%%%%%%%%%%%%%%%%%%%%%%%%%%%%%
%%%%%%%%%%%%%%%%%%%%%%%%%%%%%%%%%%%%%%%%%%%%%%%%%%%%%%%%%
\begin{figure}[ht]
\includegraphics[width=8cm,height=5cm]{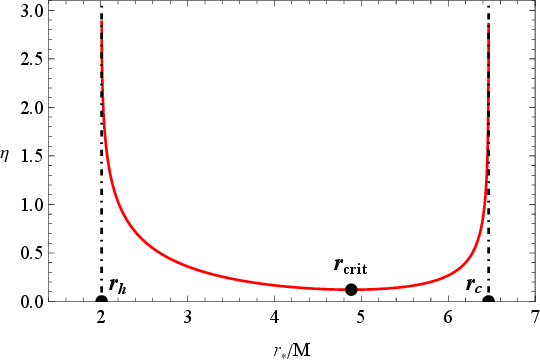}
\caption{\label{FIG14} The  efficiency $\eta$ as a function of the break up point $r_*$ for RNdS in the case $e_1>e_{crit}$. We fixed $Q=0.5M$, $\Lambda=0.05/M^2$, $m_0=2.3$, $m_1=1$, $m_2=1.1$, $q_0=2$, $q_2=8$, and $q_1=e_1=-6$. The critic charge and critic radius are $e_{crit}=-4.4056$ and $r_{crit}=4.8844$, respectively. The efficiency is positive in the whole region and $r_{crit}$ corresponds to the minimum efficiency. The event horizon and cosmological horizon are $r_h=2.0113$ and $r_c=6.4651$.}
\end{figure}
%%%%%%%%%%%%%%%%%%%%%%%%%%%%%%%%%%%%%%%%%%%%%%%%%%%%%%%%%%%%
(iii) $e_1 > e_{crit}$,  in this scenario the efficiency $\eta$ is always positive for any $r$ between the horizons, see Fig.\ \ref{FIG14}.  In the three cases  $r_{crit}$ corresponds to a minimum efficiency that surprisingly is independent of $Q$, according to Eq. (\ref{rcrit}).\\
\noi So far we have discussed the case $L_1=0$; nevertheless, Eq. (\ref{eff3}) allows $L_1 \neq 0$; in this case there are two possible efficiencies, as we discuss in Sec. \ref{twofold}.  Fig. \ref{FIG15} displays the efficiency of the energy extraction from RNdS BH showing a duplicity for $L_1 \ne 0$;  one can see that the efficiency is larger for  $\epsilon=+1$ than for $\epsilon=-1$, as in  the RN BH; while $L_1 \ne 0$ diminishes the generalized ergosphere in contrast to $L_1=0$.
%%%%%%%%%%%%%%%%%%%%%%%%%%%%%%%%%%%%%%%%%%%%%%%%%%%%%%%%
 \begin{figure}[ht]
\includegraphics[width=8cm,height=5cm]{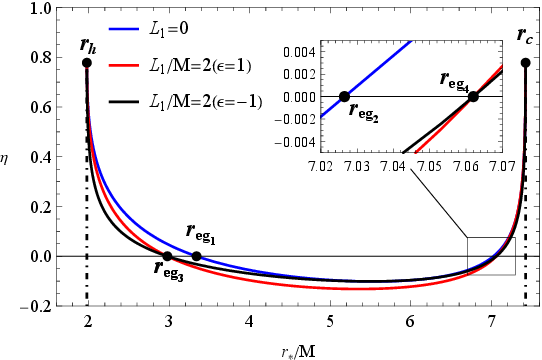}
\caption{\label{FIG15} The efficiency of the extraction process as a function of the break up point $r_*$ for RNdS BH for different values of $L_1$.  The parameters fixed as $Q=0.5M$, $m_0=2.3$, $m_1=1$, $m_2=1.1$, $q_0=4.5$, $q_1=-3.5$, $q_2=8$ and $\Lambda=0.04/M^2$. For $L_1=0$ the generalized ergospheres are $r_h<r<r_{eg_1}$ and  $r_{eg_2}<r<r_c$  and they reach  their  maxima in contrast with the case $L_1 \ne 0$. For $L_1/M=2$ there are two possible results for the efficiency $\eta$: for $\epsilon=+1$ $\eta$ is larger than for $\epsilon=-1$. The ergospheres are  defined by  $r_h<r<r_{eg_3}$ and $r_{eg_4}<r<r_{c}$. The event horizon and cosmological horizon are given by $r_h=1.97645M$ and  $r_{c}=7.425615M$, respectively. The radius of the ergospheres are $r_{eg_1}=3.33859M$ and $r_{eg_2}=7.02733M$ for $L_1=0$; $r_{eg_3}=2.97645M$ and $r_{eg_4}=7.06201M$ for $L_1=2$.}
\end{figure} 
%%%%%%%%%%%%%%%%%%%%%%%%%%%%%%%%%%%%%%%%%%%%%%%%%%%%%%%%
\subsection{Examples of the energy extraction from RNdS}
\begin{figure*}[htb]
\includegraphics[width=6cm,height=6cm]{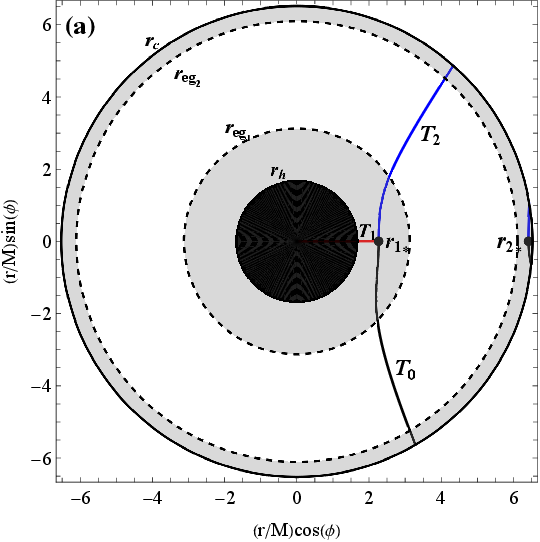}\quad
\includegraphics[width=6cm,height=6cm]{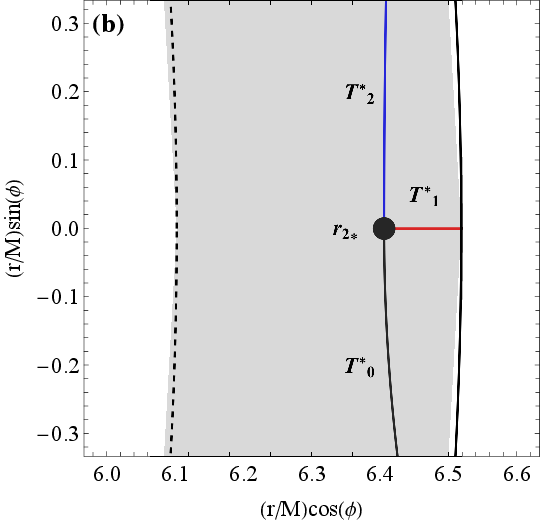}
\caption{\label{FIG16} Two examples of energy extraction from RNdS BH with the same efficiency $\eta=0.1$ for two different processes, corresponding to two different break up points $r_{1_*}=2.262794$ and $r_{2_*}=6.406015M$.  The BH parameters are $Q=0.8M$ and $\Lambda=0.05/M^2$. The outer radius of the generalized ergosphere is $r_{eg_1}=3.1222M$ and the inner radius of the cosmological ergosphere is $r_{eg_2}= 6.1027M$; the event horizon is $r_h=1.70853M$ and the cosmological horizon is $r_c=6.5188M$; the shaded areas are the ergoregions.}
\end{figure*}

Once we have described the efficiency $\eta$, in this subsection some concrete examples of the energy extraction in RNdS  illustrate two remarkably characteristics, the first one is the same resulting efficiency for different break up points $r_*$ and the second one is the possibility of a different efficiency for the same break up point $r_*$. In Fig.\ \ref{FIG16} is illustrated the extraction process from a RNdS BH; the parameters of the BH, $Q$ and $\Lambda$ are fixed and $L_1 = 0$ then the efficiency is unique, and we consider that the break up point is inside the generalized ergosphere, $r_h<r_*<r_{eg_1}$ or in the cosmological ergosphere $r_{eg_2}<r_*<r_{c}$; $T_0$ and $T_2$ describe the trajectory of the ingoing and the outgoing particle, while $T_1$ describes the trajectory of the particle with negative energy. In  Fig.\ \ref{FIG16} are shown two cases at two different break up points, where both cases have the same efficiency. Recalling that the efficiency is a ratio between the energy gained by the outgoing particle and the energy of the incident particle, then, although the efficiency is the same, the energy extracted from the BH is different, as Table\ \ref{tabla4} indicates, along the values of the rest of the parameters involved.

%%%%%%%%%%%%%%%%%%%%%%%%%%%%%%%%%%%%%%%%%%%%%%%%%%%%%
\begin{table}[htb]
\centering
\caption{Example of parameters for energy extraction from an extreme RNdS BH, characterized by $Q=0.8M$ and $\Lambda=0.05/M^2$, with an  efficiency $\eta=0.1$. The break-up point is $r_*=2.262794M$ and the radius of the ergosphere is $r_{eg_1}=3.1222M$, while the cosmological ergosphere is at $r_{eg_2}= 6.1027M$. The event horizon $r_h=1.70853M$ and the cosmological horizon $r_c=6.5188M$. These parameters generate the trajectories $T_i$ in Fig.\ \ref{FIG16}(a).} \label{tabla4}
\begin{ruledtabular}
\begin{tabular}{c c c c c c}
$i$ & $m_{i}$      & $q_{i}$  &$E_{i}$      &$L_{i}/M$&\\ \hline
0       & 2.197         & 6           & 3.12383             & 2.43878\\
1       & 1               & -2           & -0.31238           & 0\\
2      & 1               & 8          & 3.43621                & 2.43878          
\end{tabular}
\end{ruledtabular}
\end{table}

\begin{table}[htb]
\centering
\caption{A set of parameters for the  energy extraction from an extreme RNdS BH with efficiency $\eta=0.1$. The BH parameters are $Q=0.8M$ and $\Lambda=0.05/M^2$. The break-up point is $r_*=6.406015M$ and the radius of the ergosphere and cosmological horizons are $r_{eg_1}=3.1222M$ and $r_{eg_2}= 6.1027M$, respectively. The event horizon is $r_h=1.70853M$ and cosmological horizon is $r_c=6.5188M$. This set of parameters generates the trajectories $T^*_i$ in Fig.\ \ref{FIG16}(b).} \label{tabla5}
\begin{ruledtabular}
\begin{tabular}{c c c c c c}
$i$ & $m_{i}$      & $q_{i}$  &$E_{i}$      &$L_{i}/M$&\\ \hline
0       & 2.197         & 6          &1.10343             & 6.904246\\
1       & 1               & -2           & -0.11034         & 0\\
2      & 1               & 8           &1.21377               &6.904246             
\end{tabular}
\end{ruledtabular}
\end{table}

\begin{table}[H]
\centering
\caption{Example of parameters for the  energy extraction from an extreme RNdS BH with efficiency $\eta=0.0962097$. The BH  parameters are $Q=1.00892M$ and $\Lambda=0.05/M^2$. The break up point is $r_*=2M$. These parameters generate the trajectories $T_0$, $T_1$ and $T_2$ in Fig.\ \ref{FIG17}.} \label{tabla6}
\begin{ruledtabular}
\begin{tabular}{c c c c c c}
$i$ & $m_{i}$      & $q_{i}$  &$E_{i}$      &$L_{i}/M$&\\ \hline
0       & 2.3         & 6           & 4.11637            & 2.03159\\
1       & 1               & -2           & -0.396035          & 2\\
2      & 1               & 8          &4.51242               & 0.03159         
\end{tabular}
\end{ruledtabular}
\end{table}
\begin{table}[ht]
\centering
\caption{Illustration of the energy extraction from an extreme RNdS BH with efficiency $\eta=0.0784147$. The BH parameters are $Q=1.00892M$ and $\Lambda=0.05/M^2$. The break up point is $r_*=2M$. These parameters generate the trajectories $T^*_0$, $T_1$ and $T^*_2$ in Fig.\ \ref{FIG17}.} \label{tabla7}
\begin{ruledtabular}
\begin{tabular}{c c c c c c}
$i$ & $m_{i}$      & $q_{i}$  &$E_{i}$      &$L_{i}/M$&\\ \hline
$0^*$     & 2.3         & 6           &5.050522         & 8.12841\\
1            & 1                & -2         & -0.396035          & 2\\
$2^*$     & 1               & 8          &5.44656               & 6.12841          
\end{tabular}
\end{ruledtabular}
\end{table}
%%%%%%%%%%%%%%%%%%%%%%%%%%%%%%%%%%%%%%%%%%%%%%%%%%%%%%%%%%%%%%%%%%

\begin{figure}[ht]
\includegraphics[width=6cm,height=6cm]{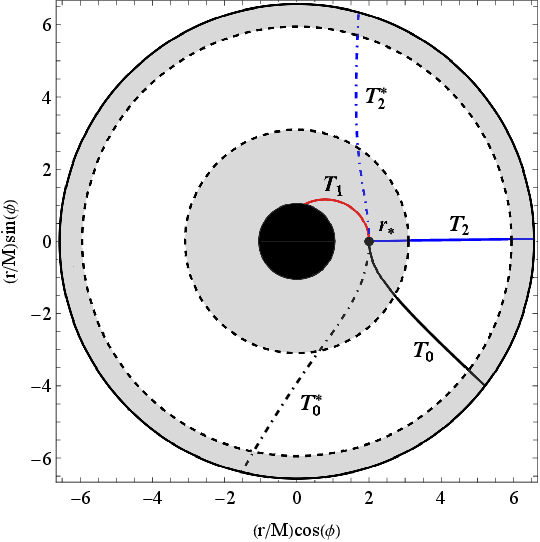}
\caption{\label{FIG17} Examples of energy extraction from RNdS BH
for the same break up point but with two different efficiencies.
The first case is described with the trajectories $T_0$, $T_1$ and $T_2$, while the second case with $T^*_0$, $T_1$ and $T^*_2$; the whole set of parameters for each process are in Table\ \ref{tabla6} and \ref{tabla7}.  The BH parameters are fixed as $Q=1.00892M$, $\Lambda=0.05/M^2$. The radius of the generalized ergosphere is $r_{eg_1}=3.09176M$ and the radius of the cosmological ergosphere is $r_{eg_2}= 5.94369M$; the event horizon is $r_h=1.03719M$ and the cosmological horizon is $r_c=6.56863M$, the shaded areas are the ergoregions.}
\end{figure}

%%%%%%%%%%%%%%%%%%%%%%%%%%%%%%%%%%%%%%%%%%%%%%%%%%%%%%%%%%%%%%%%
\noi In Fig.\ \ref{FIG17} are illustrated two different processes of extraction with the same break-up point $r_*$ for a given set of parameters ($m_0$, $m_1$, $m_2$, $q_1$, $q_2$, $L_1$); $E_2$ and $L_2$ are calculated with Eq.\ (\ref{energies}), that generates two different scenarios depending on  $\epsilon=\pm 1$, as discussed in Sec.\ \ref{twofold}. The trajectories $T_0$, $T_1$, $T_2$ correspond to $\epsilon=+1$, and $T^*_0$, $T^*_1$, $T_2$ correspond to $\epsilon=-1$. The explicit parameter values in each trajectory are indicated in Table\ \ref{tabla6} and \ref{tabla7}. 
%%%%%%%%%%%%%%%%%%%%%%%%%%%%%%%%%%%%%%%%%%%%%%%%%%%%%%%%%%%

%%%%%%%%%%%%%%%%%%%%%%%%%%%%%%%%%%%%%%%%%%%%%%%%%%%%%%%%%%%%%%%%%%%%%%%%%%%
%\vspace{0.1cm}
\section{Conclusions}
\vspace{-0.2cm}
We have analyzed in detail the process of energy extraction from the static spherically symmetric charged  BH, where the energy extraction is possible with a charged test particle (particle $0$) that splits into two particles, one of them penetrates the horizon (particle $1$) while the second one escapes to infinity (particle $2$) with a greater energy than the initial particle. We determined the energy conditions  for the particles involved in the extraction process.

We extend the study in \cite{DenardoRuffini1972} to the case of nonvanishing angular momentum $L_1 \ne 0$ for the Reissner-Nordstrom BH; in this case  the particles should satisfy $m_0>m_1+m_2$ and it turns out that the angular momentum $L_2$ and energy $E_2$ can take two possible values, consequently, there are two efficiencies for the same set of parameters ($m_0$, $m_1$, $m_2$, $q_1$, $q_2$) at a given break-up point $r_*$, see Eq. (\ref{eff3}), that allows two scenarios with different efficiencies depending on the choice of $\epsilon = \pm 1$.  For  $L_1=0$ the energy conditions in \cite{DenardoRuffini1972} are recovered.
Additionally, in Sec.\ \ref{masses} we determined the conditions that maximize the efficiency of the energy extraction process. Eq. (\ref{eff3}) defines the region where the extraction process is viable, that is called the generalized ergosphere;  this region depends on the parameters of the BH and the charge mass ratio $e$; the condition  $eQ<0$ must be fulfilled in order that the energy extraction be viable, where $Q$ is the electric charge of the BH. From Eq.(\ref{eff3}) the maximum of the efficiency depends on the location of the break-up point $r_*$ and in case it approaches the BH horizon the efficiency is  $ \eta = -q_1/q_0$.
 
The analysis of RN with negative cosmological constant, namely, the Reissner Nordstr\"om-anti-de Sitter BH (RNAdS BH) was studied in detail by \cite{Lemos2024}.
In this paper we examined the energy extraction from the Reissner-Nordstrom de Sitter black hole RNdS BH, with $\Lambda >0$. For $L_1=0$ we derive the analytic expression of the maximum radius of the generalized ergosphere, that depends on the RNdS BH parameters ($M$, $Q$, $\Lambda$) and the charge mass ratio $e$ of the test particle. In contrast with RN BH, introducing a positive cosmological constant $\Lambda>0$ generates two regions where the negative energy states (NES) are allowed and therefore the energy extraction is viable; these regions share the same characteristics as  the ones for  RN BH, namely, for a given $e$  both regions reach  their maximum size when $L_1 = 0$; also if  $e$ increases, the sizes of the ergospheres  increase as well. Also, the introduction of $\Lambda$ allows to define a critical charge $e_{crit}$ that corresponds to a critical radius $r_{crit}$ where both ergospheres join such that NES can exist throughout the whole domain of $r_h < r < r_c$; analytic expressions for $e_{crit}$ and  $r_{crit}$ are given and the cases $e_{crit} < r_{crit}$, $e_{crit} = r_{crit}$ and $e_{crit} > r_{crit}$ are illustrated. Finally, the existence of two ergospheres permits to extract energy with the same efficiency for different break-up points.   

The analysis carried out in this work applies 
not only to the Reissner-Nordstr\"om BH, but to any electrostatic and spherically symmetric BH, for instance, solutions of BH coupled to nonlinear electrodynamics  \cite{Nora2023}.

%Regarding this last issue, despite our research has dealt with an electrostatic binary metric, which in principle %cannot be assumed as a dissipative system due to its stationarity character, the study of stability of binary BHs %under external perturbations might be considered, in a similar way to the one carried out in %\cite{TesisLucimario2020} for the MP binary BH, where the BH perturbation theory \cite{Futterman} is applied to deep %insight into this matter.

%Moreover, the magnetic variant of the Penrose process that takes into account the combined influence of external %magnetic field and the rotation of a BH  seems to be connected to the origin of accretion disks where the energy %extraction into jets can befall, or even yet  the generation of ultra-high energy cosmic rays \cite{Kolos2021, %Kolos2020}.  In this direction we aim to develop further research of the magnetic Penrose process in BH binaries.
%%%%%%%%%%%%%%%%%%%%%%%%%%%%%%%%%%%%%%%%%%%%%%%%%%%%%%%%%%
\vspace{-0.5cm}
\section{Acknowledgments}
\vspace{-0.2cm}
NB acknowledges partial support by CONAHCyT Project CBF2023-2024-811. ICM acknowledges financial support of SNI-CONAHCyT, Mexico, grant CVU No. 173252. AB acknowledges financial support by CONAHCyT, Mexico, through the PhD Scholarship 814092.
%%%%%%%%%%%%%%%%%%%%%%%%%%%%%%%%%%%%%%%%%%%%%%%%%%%%%%%%%%%%%%%%%%%

\end{document}